\documentclass[12pt]{article}
\usepackage{amsfonts}
\usepackage{amsmath,amssymb}
\usepackage{array}
\usepackage{color}
\usepackage{amsmath}
\usepackage{amsxtra}
\usepackage{amstext}
\usepackage{amssymb}
\usepackage{mathrsfs}
\usepackage{latexsym}
\newtheorem{thm}{Theorem}[section]

\newtheorem{lem}[thm]{Lemma}
\newtheorem{remark}[thm]{Remark}

\newtheorem{definition}[thm]{Definition}

\numberwithin{equation}{section}
\usepackage{color}
\newcommand{\bl}{\textcolor{blue}}
\newcommand{\rd}{\textcolor{red}}

\def\Tr{{\rm Tr\,}}

\def\Det{{\rm Det}}
\def\R{\Bbb R}
\def\C{\Bbb C}

\def\N{\Bbb N}

\def\supp{{\rm supp\,}}

\def\Det{{\rm Det\,}}
\newcommand{\thedate}

\begin{document}

\phantom{.} {\qquad \hfill \textbf{\bl{MPRF}}}

\bigskip

\bl{\textbf{Dedicated to Robert A. MINLOS in occasion of his 80th birthday}}

\vskip 2.5cm

\begin{center}

{\Large{\bf {Random point field approach to analysis of}}}

{\Large{\bf {anisotropic Bose-Einstein condensations }}}

\vskip 1cm

\setcounter{footnote}{0}
\renewcommand{\thefootnote}{\arabic{footnote}}

{\textbf{Hiroshi Tamura} \footnote{tamurah@kenroku.kanazawa-u.ac.jp}\\
       Graduate School of the Natural Science and Technology\\
           Kanazawa University,\\
          Kanazawa 920-1192, Japan

\bigskip

\textbf{Valentin A.Zagrebnov }\footnote{Valentin.Zagrebnov@univ-amu.fr}\\
D\'{e}partement de Math\'{e}matiques - Universit\'{e} d'Aix-Marseille (AMU)\\
Centre de Physique Th\'eorique - UMR 7332 \\
Luminy-Case 907, 13288 Marseille Cedex 9, France}

\vspace{1.5cm}

\end{center}

\begin{abstract}

Position distributions of constituent particles of the perfect Bose-gas
trapped in exponentially and polynomially anisotropic
boxes are investigated by means of the boson random point fields (processes)
and by the spatial random distribution of particle density.
Our results include the case of \textit{generalised} Bose-Einstein Condensation.

For \textit{exponentially} anisotropic quasi two-dimensional system (SLAB),
we obtain \textit{three} qualitatively different particle density
distributions. They correspond to the \textit{normal} phase,
the \textit{quasi-condensate} phase (\textit{type III} generalised
condensation) and to the phase when the \textit{type III} and the
\textit{type I} Bose condensations \textit{coexist}.

An interesting feature is manifested by the \textit{type II}
generalised condensation in one-directional polynomially anisotropic system (BEAM).
In this case the particle density distribution rests \textit{truly}
random even in the \textit{macroscopic} scaling limit.

\end{abstract}

\vspace{1.5cm}

\noindent \textbf{Key words:} random point fields (processus); quantum statistical mechanics;
Bose-Einstein condensation; anisotropic systems; spacial particle density.

\tableofcontents

\vspace{1.5cm}

\section{Introduction and Main Results}
The microscopic position distribution of constituent particles of
quantum gases in statistical mechanics can be described by Random Point
Fields (RPFs, or Random Point Processes). For bosons this idea
goes back at least to \cite{F91, FF91}. \textit{Determinantal} and \textit{permanental} processes describe, respectively,
the position distributions of ideal Fermi-and Bose-gases \cite{M75, Ly}. It is known that position
distribution of bosons in the state of Bose-Einstein Condensation (BEC) can be described by the
corresponding \textit{permanental} RPF explicitly \cite{2, E, TZa}.
For a introduction into the theory of general RPFs as well as the fermion- and the boson-RPF and their applications
in Statistical Mechanics, see, e.g., \cite{DV, GeoY, ST03, So, 1, 3}.

It is also possible to prove the Large Deviation Principle (LDP) for
the boson RPFs in the condensed and non-condensed phases \cite{ST03, TZb}, extending some known results in this field \cite
{BruZ}, although the critical point needs a separate analysis.The proof of the LDP involves large-scale test functions
for the generating functional of the boson RPF, which one must scale properly  for different phases, see discussion in
\cite{TZb}.
Alternatively, the large-scale behavior of position distribution, or macroscopic (or \textit{mesoscopic}) position
distributions,can be treated as a \textit{particle density}, which can be described by Random Fields (RFs). For example,
RF which represents macroscopic boson density is obtained by scaling the RPF for boson gas trapped by external potential,
see \cite{TZa, T}. However, in the last case the RF yields a deterministic density distribution in the semiclassical limit,
which is similar to the law of large numbers \cite{TZb}.

Recently, the study of boson gases trapped in quasi one-dimensional and quasi two-dimensional regions have been developed
\cite{BZ} with the aim to elucidate physical arguments and experiments, which indicate the
existence of \textit{generalised} (gBEC), or \textit{quasi}-BEC (qBEC), boson gases trapped in extremely anisotropic systems
\cite{KvD, vDK, G, H}. The unusual properties of highly anisotropic boson systems are known since eighties and they can be
related to the notion of \textit{generalized} BEC by van den Berg-Lewis-Pul\'{e} (vdBLP-gBEC) \cite{BLP}.
Although their observation and classification of the gBEC in types I, II and III seems purely mathematical at the first glance,
it is accepted to be quite relevant to the interpretation of the experimental data in
extremely anisotropic systems, such as SLABs, BEAMs and CIGARs, see discussion in \cite{BZ} and the recent paper \cite{MuSa}.

However the case of the anisotropic CIGAR shape, which is the most interesting for traps, is not covered by the van den
Berg-Lewis-Pul\'{e} theory. For the first time this case was rigorously studied in \cite{BZ}.
It gives a qualitative description of the experiments \cite{vDK, G, H},
including the existence of \textit{two} critical points and \textit{transition} between type-I and type-III condensates, i.e.
between normal conventional BEC and "quasi-condensation". The peculiar property of the CIGAR systems seems to come from their
exponential one-direction anisotropy.
Although the BEAM anisotropy may produce the type III condensation, it does not split the \textit{single} critical point into
\textit{two} points, if one does not modify external potential in the anisotropic direction \cite{BZ}.

The aim of the present paper is to study the features of the generalised BEC phases from the view point of position
distributions. To this end we consider RFs which describe scaled particle density
as well as RPF for BEC in two cases: the \textit{exponentially} anisotropic SLAB and \textit{algebraically} anisotropic
BEAM which are obtained by the thermodynamic limit of Casimir prisms.

As mentioned above, the SLAB system manifests three phases: normal phase (without BEC), condensation of type-III and
coexistence of condensations type-I and type-III, separated by \textit{two} critical points.
So, first we construct the RPF, which describes the microscopic position
distribution of constituent bosons, for these three domains. This result clearly demonstrate the difference between
the normal and the BEC
phases. However, the qualitative difference between two BEC phases (type-III and the mixture of type-III and type-I)
becomes evident only after contraction to the particle density, obtained RPs by scaling the boson RPF.
We find that the density distributions are essentially deterministic,  i.e., the corresponding RFs are non-random.
This result is similar to the behaviour of the particle density
distribution in the ``semi-classical" limit for the boson mean-field model in a weak harmonic trap \cite{T}.

In contrast to the SLAB model, the BEAM is an \textit{algebraically} anisotropic case of Casimir prisms.
A specific choice of anisotropy rate gives BEAM a stronger randomness
to the particle density distribution than the SLAB. It is known that anisotropic BEAM has only \textit{one} critical point
\cite{BLP,BZ}. Our specific choice of the anisotropy rate (see Section \rd{1.3}) implies that this point separates the normal
phase and the type-II condensation. Since the type-II condensation is in sense a type-I condensation smeared out over a
countable infinite number of states \cite{BLP,ZB}, it is expected that the particle number fluctuations in type-II condensate
should be higher than the type-III condensate phase.\cite{BLP,BLL}
Our result shows that for this kind of BEAM, the \textit{macroscopic} particle density RF obtained by scaling
(similar to the SLAB case) from RPF gives non-trivial fluctuations described by squared gaussian distribution.

In the next subsection, a brief introduction to boson RPF is presented
in order to describe the position distribution of the free boson gas in three dimensional anisotropic boxes.
In subsection 1.2, we formulate  theorems on RPFs and RFs concerning
micro-, macro- and mesoscopic behaviour of position distribution of
the boson gas in exponentially anisotropic SLAB as well as their
corollaries. In subsection 1.3, we collect the corresponding results about \textit{algebraically} anisotropic
(in one direction) Casimir prisms, that for a spacial rate converges to the BEAM.
Section 2 and 3 are devoted to proofs of the theorems for SLAB and for BEAM, respectively.

\subsection{Preliminaries: Notations and Definitions}
\textbf{(a) Random Point Fields (Processes).}
Let $E$ be a locally compact metric space serving as the state-space of the \textit{point}
configurations $\xi \subset E$. By $\mathfrak{B}$ we denote the corresponding Borel $\sigma$-algebra on $E$
and by $\mathfrak{B}_0 \subseteq \mathfrak{B}$ the \textit{relatively} compact Borel sets in $E$.
We denote by $\mu$ a \textit{diffusive} (i.e. $\mu(x)=0$ for any one-element subset $x\in E$) locally finite reference measure
on $(E, \mathfrak{B})$. (The standard example is the Lebesgue measure $\mu(dx) = dx$ on $(E = \mathbb{R}^d, \mathfrak{B})$.)

We denote by $Q_{E}$ the subspace of \textit{locally-finite} point configurations $\{\xi\subset E\}$ :
\begin{equation*}
Q_{E} := \{\xi \subset E:\ {\rm{{card}}} (\xi \cap \Lambda) < \infty \ {\rm{\ for \ all}}\  \Lambda \in \mathfrak{B}_0 \} \ .
\end{equation*}
Hence, for any $\Lambda \in \mathfrak{B}_0$ one can define a subspace of the point configurations
$Q_{\Lambda}:= \{\xi \in Q_{E}: \xi \subset \Lambda \}$ and the mapping $\pi_{\Lambda}: \xi \mapsto \xi \cap \Lambda$ for
the corresponding projection from $Q_{E}$ onto $Q_{\Lambda}$. Then \textit{counting} function:
$N_\Lambda : \xi \mapsto {\rm{{card}}} (\pi_{\Lambda}(\xi))$ is finite for any $\Lambda \in \mathfrak{B}_0$.

Now one can introduce the notion of the \textit{spatial} random point field (RPF) on $\mathbb{R}^d$ as locally finite
discrete random sets $\xi \subset \mathbb{R}^d$, i.e. such that $N_\Lambda(\xi) < \infty$ for $\Lambda \in \mathfrak{B}_0$.
Since below we use the Laplace transformation for characterisation of the RPFs, we need a more elaborated general setting.

Let $\delta_{x}$ denote the atomic measure on $\mathfrak{B}$ supported at one-element subset $x\in E$. Then
any configuration of points $\xi \in Q_{E}$ can be \textit{identified} with the non-negative integer-valued  {Radon measure}:
$\lambda_\xi (\cdot) := \sum_{\{x\in \xi\}} \delta_x (\cdot)$ on the Borel $\sigma$-algebra $\mathfrak{B}$. Hence,
$\lambda_{\xi}(D)= N_D(\xi)$ is the number of points that fall into the set $D\in\mathfrak{B}_0$ for the locally finite
point configuration $\xi \in Q_{E}$. Recall that $C_{0}(E)^{\ast}$, which is dual to the space of continuous on $E$ functions
$C_{0}(E)$ vanishing at infinity and equipped with the uniform norm, is isometric (by the Riesz representation theorem) to
the space $\mathcal{M}(E)$ of Radon measures on $E$. By this isometry the {weak-$\ast$} topology on $C_{0}(E)^{\ast}$ yields
the \textit{vague} topology on $\mathcal{M}(E)$. Then identification of $\mathcal{M}(E)$ with the set of Radon
measures $\lambda_\xi$ induces on the point configuration space $Q_{E}$ a topology, turning $Q_{E}$ into a locally
compact separable metric space with the corresponding Borel $\sigma$-algebra $\mathfrak{B}(Q_{E})$. Note that if
$\mathfrak{F}$ is the smallest $\sigma$-algebra on $Q_{E}$ such that the mappings $N_\Lambda$ are measurable for all
$\Lambda \in \mathfrak{B}_0$, then $\mathfrak{F} = \mathfrak{B}(Q_{E})$, see e.g. \cite{DV}.
\begin{definition}\label{def-RPP1} A random point field (process) is a triplet $(Q_{E}, \mathfrak{B}(Q_{E}), \nu)$, where
$\nu$ is a probability measure on $(Q_{E}, \mathfrak{B}(Q_{E}))$. Its marginal on $Q_{\Lambda}$ is defined by the probability
measure $\nu_{\Lambda}:= \nu \circ \pi_{\Lambda}^{-1}$.
\end{definition}

Note that the process defined above is \textit{simple}, i.e. the random measure $\lambda_\xi $ almost surely assigns measure
$\leq 1$ to \textit{singletons}.\\
\textbf{(b) Correlation Functions and Laplace Transformation.}
For the marginal measure $\nu_{\Lambda}$ we consider the \textit{Janossy} probability densities
$\{j_{\Lambda,s}(x_1, \ldots , x_s)\}_{s\geq 0}$, \cite{DV}. Here
$j_{\Lambda,s=0}(\emptyset) = \nu_{\Lambda}(\{\xi: N_{\Lambda}(\xi)=0\})$
and for $s \geq 1$ it is a joint probability distribution that there are \textit{exactly} $s$ points in $\Lambda$, each
located in the vicinity of the one of $x_1, \ldots , x_s$, and no points elsewhere. By construction the Janossy probability
densities are \textit{symmetric} and verify the \textit{normalization} condition
\begin{equation}\label{norm-cond-Janossy}
\sum_{s=0}^{\infty} \frac{1}{s !} \ \int_{\Lambda^{s}} \ \mu(dx_1) \ldots \mu(dx_s) \ j_{\Lambda,s}(x_1, \ldots , x_s) = 1 \ ,
\end{equation}
with a standard \textit{convention} for $s=0$.
Then for any measurable function $F$ on $Q_{\Lambda}$ with components $\{F_s\}_{s\geq0}$ one gets \cite{DV}:
\begin{eqnarray}\label{RPF-Janossy}
\int_{Q_{\Lambda}} \nu_{\Lambda}(d \xi) F(\xi)=
\sum_{s=0}^{\infty} \frac{1}{s !} \ \int_{\Lambda^{s}} \ \mu(dx_1) \ldots \mu(dx_s) \ j_{\Lambda,s}(x_1, \ldots , x_s)\
F_{s}(x_1, \ldots , x_s) \ .
\end{eqnarray}

These joint probability distributions (\textit{correlation functions}) serve for a very useful characterization of RPFs by
Laplace transformation. Let $f: E \rightarrow \mathbb{R}_{+}$, be non-negative continuous function with compact support.
For each $f$ one can define by
\begin{equation}\label{def-funct-onE}
\langle f,\xi\rangle := \int_{E} \lambda_\xi (d x) f(x)= \sum_{x\in \xi} f(x) \ ,
\end{equation}
the measurable function: $\xi \mapsto \langle f,\xi\rangle$ on $Q_{E}$. Then by virtue of (\ref{RPF-Janossy}) the Laplace
transformation of the measure $\nu_{\Lambda}$ for a given $f$ takes the form
\begin{eqnarray}\label{Lapl-trans}
&&\mathbb{E}_{\nu_{\Lambda}}(e^{-\langle f,\xi\rangle }) = \int_{Q_{\Lambda}} \nu_{\Lambda} (d \xi) \
e^{-\langle f,\xi\rangle } = \\
&&\sum_{s=0}^{\infty} \frac{1}{s !} \ \int_{\Lambda^{s}} \ \mu(dx_1) \ldots \mu(dx_s) \ j_{\Lambda,s}(x_1, \ldots , x_s)\
\prod_{j=1}^{s} e^{- f(x_j)} \ . \nonumber
\end{eqnarray}

The most fundamental example of RPF is the Poisson point process $\pi_{z}(d \xi)$ on $E = \mathbb{R}^d$ with Lebesgue
measure $\mu(dx) =  dx$ and the \textit{intensity} function $z(x)\geq 0$. For this RPF its marginal on $Q_\Lambda$ is
defined by the {Janossy} probability densities:
\begin{equation*}
\{j_{\Lambda,s}(x_1, \ldots , x_s) = e^{- \int_{\Lambda} dx \, z(x)} \ \prod_{j=1}^{s} z(x_j)\}_{s\geq 0} \ .
\end{equation*}
Then taking into account (\ref{Lapl-trans}) one gets for any non-negative continuous function $f$ with compact support
the corresponding Laplace transformation (\textit{generating} functional) expressed by the well-known formula:
\begin{equation}\label{Lapl-Poisson}
\mathbb{E}_{\pi_{z}}(e^{-\langle f,\xi\rangle })= \exp\left\{- \int_{\mathbb{R}^d} dx \, z(x) (1 - e^{-f(x)}) \right\} \ ,
\end{equation}
for extension to infinite configurations $Q_{\mathbb{R}^d}$.\\
\textbf{(c) Permanental (boson) RPF.}
In the box $ \Lambda = \prod_{j=1}^3[-L_j/2, L_j/2] \subset \mathbb{R}^3 $,
we consider a quantum mechanical system of identical free bosons of the mass $m$ with the one-particle Hamiltonian
\begin{equation}\label{1-partHam}
H_{\Lambda} = - \frac{\hbar^2}{2m}\triangle_D \ ,
\end{equation}
where $\triangle_D$ denotes the Laplacian operator with Dirichlet boundary conditions in the space $L^2(\Lambda)$ .
The spectrum and the eigenfunctions of this self-adjoint operator are given by
\begin{equation}\label{symm-spectr}
       \bigg\{ \epsilon_\mathbf{k} = \frac{\hbar^2}{2m}\sum_{j=1}^3
          \Big( \frac{\pi k_j}{L_j} \Big)^2 \, \bigg| \,
         \mathbf{k} = (k_1, k_2, k_3) \in \N^3 \bigg\} \ ,
\end{equation}
\begin{equation}\label{symm-eigenfunct}
\bigg\{ \phi_{\mathbf{k}, \Lambda}(x) = \prod_{j=1}^3\sqrt{\frac{2}{L_j}}
           \sin\Big(\frac{\pi k_j}{2} + \frac{\pi k_j x_j}{L_j}\Big)\bigg| \,
         \mathbf{k} \in \N^3 \bigg\} \ .
\end{equation}
Then the corresponding trace-class valued (Gibbs) semigroup \cite{Zag} and its kernel are
\begin{equation}\label{GibbS1}
\left\{G_{\Lambda}(\beta):= e^{-\beta H_{\Lambda}}\in \mathcal{C}_{1}(L^2(\Lambda))\right\}_{\beta \geq0}\,, \
G_{\Lambda}(\beta; x, y):= \sum_{\mathbf{k}\in \N^3}e^{-\beta\epsilon_\mathbf{k}}
\phi_{\mathbf{k}, \Lambda}(x)\overline{\phi_{\mathbf{k}, \Lambda}(y)} \ .
\end{equation}

The random point field defined by this kernel is the Boson (BRPF), i.e. the \textit{permanental} RPF \cite{DV}. It
describes, in correspondence with the RPF paradigm, the \textit{position} distribution of non-interacting bosons in the
box $\Lambda$. If the Bose-gas is in the \textit{grand-canonical} equilibrium state at fixed temperature $\beta^{-1}$
and chemical potential $\mu$, then the BRPF generating functional (cf.(\ref{Lapl-Poisson})) has the form \cite{TZa}, \cite{T}:
\begin{equation}\label{gf0}
\mathcal{Z}_{\Lambda, \mu}^{gc}[f] = \int _{Q_{\Lambda}}  \nu_{\Lambda, \mu}(d\xi) \ e^{-\langle f, \xi \rangle}
= \frac{\Xi_{\Lambda, f}(\mu)}{\Xi_{\Lambda, 0}(\mu)} \ \ .
\end{equation}
Here denominator is the perfect Bose-gas \textit{grand-canonical} partition function for $\mu \leq 0$:
\begin{eqnarray}
\Xi_{\Lambda, 0}(\mu) &:=& \sum_{n=0}^{\infty}\int_{\Lambda^n} \, dx_{1} \ldots dx_{n}\
\frac{e^{n \beta \mu}}{n!} {\rm{Per}}\{G_{\Lambda}(\beta; x_{i},x_{j})\}_{1\leqslant i,j \leqslant n} \nonumber \\
&=&\Det[1-e^{\beta \mu}G_{\Lambda}]^{-1} \ ,
\label{GC-PartFunc}
\end{eqnarray}
and the numerator
\begin{eqnarray}
\Xi_{\Lambda, f}(\mu) &:=& \sum_{n=0}^{\infty}\int_{\Lambda^n} \, dx_{1} \ldots dx_{n} \
\frac{e^{n \beta \mu}}{n!} \, e^{-\sum_{s=1}^nf(x_{s})}
\, {\rm{Per}}\{G_{\Lambda}(\beta; x_{i},x_{j})\}_{1\leqslant i,j \leqslant n} = \nonumber \\
&=& \Det[1-e^{\beta \mu} e^{-f} G_{\Lambda}]^{-1} =
\Det[1-e^{\beta \mu} G_{\Lambda}^{1/2} e^{-f} G_{\Lambda}^{1/2}]^{-1} \ ,
\label{GC-PartFunc-f}
\end{eqnarray}
where Det denotes the Fredholm \textit{determinant} and Per stands for the \textit{permanent} of the
corresponding operators. The both representations (\ref{GC-PartFunc}) and (\ref{GC-PartFunc-f}) via determinants are due to
Vere-Jones' formula generalized to the trace-class operators $G_{\Lambda}$ in Theorem 2.4, \cite{ST03}.

Notice that in the \textit{grand-canonical} ensemble the BRPF configurations $Q_{\Lambda}$ are \textit{infinite}. To consider
in $\Lambda$ the finite point BRPF $\nu_{\Lambda, n}(d\xi)$ one has to consider the \textit{canonical} ensemble \cite{1,2}.
Then the probability density of the $n$-point configurations in $\Lambda$ is defined by
\begin{equation}\label{CE-1}
p_{\Lambda,n} (x_1, \ldots , x_n) = \frac{1}{n! \ {Z}_{\Lambda, 0}(n)}
{\rm{Per}}\{G_{\Lambda}(\beta; x_{i},x_{j})\}_{1\leqslant i,j \leqslant n} \ ,
\end{equation}
where
\begin{equation}\label{CanonPart}
{Z}_{\Lambda, 0}(n):= {\rm{Tr}}_{\, \mathcal{H}_{n}} \, e^{- \beta H_{\Lambda,n}}=\int_{\Lambda^n} \, dx_{1} \ldots dx_{n}\
\frac{1}{n!} \ {\rm{Per}}\{G_{\Lambda}(\beta; x_{i},x_{j})\}_{1\leqslant i,j \leqslant n} \ ,
\end{equation}
is the $n$-boson \textit{canonical} partition function defined by the trace over the Hilbert space
$\mathcal{H}_{n}:= L_{sym}^2(\Lambda^n)$ of symmetrized function \cite{1}. Here the sum
\begin{equation*}
H_{\Lambda,n}: = \sum_{1\leq j \leq n} (- \frac{\hbar^2}{2m} \triangle_{j,D} ) \ ,
\end{equation*}
defines the $n$-particle Hamiltonian in the space $L^2(\Lambda^n)$, cf. (\ref{1-partHam}). Similarly to (\ref{GC-PartFunc-f})
we also define
\begin{equation}\label{CanonPart-f}
{Z}_{\Lambda, f}(n):=\int_{\Lambda^n} \, dx_{1} \ldots dx_{n}\
\frac{1}{n!} \ e^{-\sum_{s=1}^nf(x_{s})}\, {\rm{Per}}\{G_{\Lambda}(\beta; x_{i},x_{j})\}_{1\leqslant i,j \leqslant n} \ .
\end{equation}

Then according to construction presented in \textbf{(a)}, the $n$-point BRPF $\mu_{\Lambda, n}(d\xi)$ in $\Lambda$ is
induced by the map $(x_1, \ldots , x_n) \mapsto \sum_{j=1}^{n} \delta_{x_j} (\cdot) \in Q_{\Lambda, n}$ and probability
measure on $\Lambda^n$ with density (\ref{CE-1}). The BRPF generating functional in canonical ensemble has the form
\cite{1,2}:
\begin{equation}\label{CE-2}
\mathcal{Z}_{\Lambda, n}[f] = \int _{Q_{\Lambda, n}}  \mu_{\Lambda, n}(d\xi) \ e^{-\langle f, \xi \rangle}
= \frac{Z_{\Lambda, f}(n)}{Z_{\Lambda, 0}(n)} \ \ ,
\end{equation}

Note that by (\ref{RPF-Janossy}) and (\ref{Lapl-trans}) the Janossy probability densities of the \textit{finite-volume}
BRPF for a fixed grand-canonical temperature $\beta$ and the chemical potential $\mu$ are
\begin{equation}\label{BRPF-Janossy}
j_{\Lambda,n}(\beta,\mu; x_1, \ldots , x_n):= \frac{1}{\Xi_{\Lambda}(\mu)} \  e^{n \beta \mu} \
{\rm{Per}}\{G_{\Lambda}(\beta; x_{i},x_{j})\}_{1\leqslant i,j \leqslant n}  \ ,
\end{equation}
see (\ref{gf0}). Here $\{n= 0,1, \ldots\}$ and ${\rm{Per}}\{G_{\Lambda}(\beta; x_{i},x_{j})\}_{1\leqslant i,j
\leqslant n =0} = 1$.

Since the function $f$ is supposed to be \textit{non-negative continuous} with a \textit{compact} support, the
fundamental properties of the (Fredholm) determinant and (\ref{gf0}) imply that:
\begin{equation}\label{gfD}
\mathcal{Z}_{\Lambda, \mu}^{gc}[f] = \Det[1 + \sqrt{1-e^{-f}}K_{\Lambda}(z)\sqrt{1-e^{-f}}]^{-1} \ , \ \
\ z:= e^{\beta \mu} \ .
\end{equation}
Here for $\mu <0$, the positive operator $K_{\Lambda}(z): = z G_{\Lambda}(1- zG_{\Lambda})^{-1} > 0$, is self-adjoint
and belongs to the trace-class $\mathcal{C}_{1}(L^2(\Lambda))$ of operators on $L^2(\Lambda)$.
We also denote by
\begin{equation}\label{Kf-z}
K_{\Lambda,f}(z) : = \sqrt{1-e^{-f}}K_{\Lambda}(z)\sqrt{1-e^{-f}} \ \in \ \mathcal{C}_{1}(L^2(\Lambda)) \ ,
\end{equation}
the trace-class operator in determinant (\ref{gfD}).

In order to make the generating functional $\mathcal{Z}_{\Lambda, \mu}^{gc}[f]$ well-defined, i.e. to ensure the
convergence in the numerator and the denominator (\textit{stability}), the chemical potential $\mu$ \textit{must} satisfy
the condition $e^{\beta\mu}G_{\Lambda} < 1$, i.e.,$ \mu < \epsilon_{\mathbf{1}} = \sum_{j=1}^3{\pi^2 \hbar^2}/{2m}{L_j}^2 $.
Here ${\bf{1}} = (1,1,1) \in \N^3$.

Hence, from now on we put $\mu: = \epsilon_{{\bf{1}}= (1,1,1)} - \Delta$, where for stability we assume that $\Delta > 0$.
Since below we examine the behaviour of BEC in the thermodynamic limit $\{L\rightarrow\infty:L_1, L_2, L_3 \to \infty\}$
for \textit{anisotropic} vessels, we assume that $\Delta = \Delta_L$ may depend on $L_1, L_2, L_3$ and that the value of
$ (L_1 L_2 L_3) \, \Delta_L$ is bounded. The latter ensures \textit{boundedness} of the total particle density:
\begin{equation}\label{meanPartDens}
\rho  =  \frac{1}{L_1 L_2 L_3} \
\sum_{\textbf{k} \in \mathbb{N}^3} \frac{1} {\exp \beta [({\hbar^2\pi^2}/{2m})\sum_{j=1}^3 {(k_j^2-1)}/{L_j^2} +
\Delta_L ]-1} \ ,
\end{equation}
in the {grand-canonical} ensemble $(\beta, \mu)$, including the limit ${L\rightarrow\infty}$. Here
$\{\Delta_L = \Delta_L (\beta,\rho)\}_L$ is solution of the equation (\ref{meanPartDens}), \cite{ZB}, for a fixed bounded
density $\rho$.

Notice that the integral kernel of the operator $K_{\Lambda}$ in the grand-canonical ensemble $(\beta, \mu =
\epsilon_{\bf{1}} - \Delta)$ is given by
\begin{equation}\label{ODLROkernel}
K_{\Lambda}(x, y):= \sum_{\textbf{k}\in\N^3}
\frac{\phi_{\textbf{k}, \Lambda}(x)\overline{\phi_{\textbf{k}, \Lambda}(y)}}
{\exp \beta [({\hbar^2\pi^2}/{2m})\sum_{j=1}^3
{(k_j^2-1)}/{L_j^2} + \Delta ]-1} \ .
\end{equation}
Since the series in the right-hand side of (\ref{ODLROkernel}) is uniformly convergent, the kernel is continuous.\\
\textbf{(d) Equivalence of ensembles and the Kac probability distribution.}
To make a contact between canonical $(\beta,\rho)$ grand-canonical $(\beta,\mu)$ ensembles we use to the generating functional
(\ref{gf0}) for the case of $f(x):= t \, \mathbb{I}_{\Lambda}(x)/|\Lambda|$,  $t \geq0$. Here $\mathbb{I}_{\Lambda}(x)$
is indicator of domain $\Lambda$ with volume which is $V:= |\Lambda|$. Then by virtue of (\ref{def-funct-onE}) the functional
(\ref{gf0}) takes the form
\begin{equation}\label{Kac1}
\mathbb{E}_{\beta, \, \mu}(e^{- t \, N_{\Lambda}(\xi)/V}) =
\frac{\Xi_{\Lambda}(\beta, \mu - t/(\beta V))}{\Xi_{\Lambda}(\beta, \mu)} =
\prod_{\textbf{k}\in \mathbb{N}^3 } \frac{1- e^{- \beta(\epsilon_{\textbf{k}} - \mu)}}
{1- e^{- \beta(\epsilon_{\textbf{k}} - \mu + t/(\beta V))}}\ ,
\end{equation}
where $\mathbb{E}_{\beta, \mu}(\cdot)$ is expectation in the Gibbs grand-canonical ensemble $(\beta,\mu)$ and
$\Xi_{\Lambda}(\beta, \mu) \equiv \Xi_{\Lambda, 0}(\mu)$ (\ref{GC-PartFunc}).
Notice that (\ref{Kac1}) has also representation:
\begin{equation}\label{Kac2}
\mathbb{E}_{\beta, \, \mu}(e^{- t \, N_{\Lambda}(\xi)/V})=
\sum_{n=0}^{\infty}  \ \frac{e^{\beta n \mu }\ {Z}_{\Lambda}(\beta, n)}{\Xi_{\Lambda}(\beta, \mu)} \ e^{- t n/V} \ ,
\end{equation}
where $Z_{\Lambda}(\beta, n)\equiv Z_{\Lambda,0}(n)$ is defined by (\ref{CanonPart}).

For domain out of condensation, i.e. for $\mu<0$, the thermodynamic limit of (\ref{Kac2}) follows directly from representation
(\ref{Kac1}):
\begin{equation}\label{Kac3}
\lim_{V\rightarrow\infty}\mathbb{E}_{\beta,\, \mu}(e^{- t \, N_{\Lambda}(\xi)/V})=
\int_{0}^{\infty} dx \, \delta_{\rho(\beta,\mu)} (x) \ e^{-t x}  = e^{-t \rho(\beta,\mu)}\ .
\end{equation}
Here
\begin{equation}\label{rho<rho-c}
\rho(\beta,\mu) :=  \frac{1}{(2 \pi)^3}\int_{\mathbb{R}^3}\frac{d^3 k}{e^{\beta(\epsilon_{\textbf{k}} - \mu)}-1} \ ,
\end{equation}
is the particle grand-canonical density for $\mu \leq 0$, which reaches its maximal (critical) value at $\mu = 0$:
\begin{equation}\label{rho-crit}
\rho_{c}(\beta):=\rho(\beta,\mu =0)
\end{equation}
Note that (\ref{Kac3}) expresses the \textit{strong equivalence} of ensembles \cite{ZP} for the perfect Bose-gas in this
one phase regime:
\begin{equation}\label{Eq-Ens}
\mathbb{E}_{\beta, \, \mu}(A)=
\int_{0}^{\infty} dx \, \delta_{\rho(\beta,\mu)}(x) \ \mathbb{E}_{\beta,\, x}(\hat{A}) =
\mathbb{E}_{\beta, \, \rho(\beta,\mu)}(\hat{A}) \ ,
\end{equation}
where $\mathbb{E}_{\beta, x}(\cdot)$ is the canonical quantum Gibbs state for the total particle density $x$, and $A$,
$\hat{A}$ are corresponding observable in grand- and in canonical ensembles \cite{BNZ}.

Hence, for $\rho < \rho_{c}(\beta)$ the solution of equation (\ref{meanPartDens}) gives $\mu_{L}(\beta, \rho)$ such that
the limit $\lim_{V\rightarrow\infty}\mu_{L}(\beta, \rho)= {\mu}(\beta, \rho) < 0$, and $\rho(\beta, {\mu}(\beta, \rho)) =
\rho$ by (\ref{rho<rho-c}).
\begin{definition}\label{mu-def} We define the chemical potential $\overline{\mu}:=\overline{\mu}(\beta,\rho)$ as the inverse
function of density (\ref{rho<rho-c}), (\ref{rho-crit}), with extension: $\overline{\mu}(\beta,\rho \geq \rho_{c}(\beta)):= 0$.
Hence,
$\rho(\beta, \overline{\mu}(\beta, \rho)) = \rho$ and $\rho(\beta, \overline{\mu}(\beta, \rho \geq \rho_{c}(\beta))) =
\rho_{c}(\beta)$.
\end{definition}

The phase with condensation corresponds to $\rho > \rho_{c}(\beta)$, i.e. $\lim_{V\rightarrow\infty}\mu_{L}(\beta, \rho)= 0$.
Now to deduce the asymptotics of the solution $\mu_{L}(\beta, \rho)$ one has to separate from the sum (\ref{meanPartDens})
the most singular term ${\bf{1}} = (1,1,1)$. Let $\mu_{L}(\beta, \rho) := \epsilon_{1} - \Delta_L (\beta,\rho)$.
Then for thermodynamic limit (in the Fisher sense \cite{Ru}) it has the form:
\begin{equation}\label{rho>rho-c}
\mu_{L}(\beta, \rho) := \epsilon_{1} - \frac{1}{\beta (\rho - \rho_{c}(\beta)) V} + o(V^{-1}) \ .
\end{equation}
The asymptotics (\ref{rho>rho-c}) ensures that the single term ${\bf{1}} = (1,1,1)$ gives particle density
$\rho_{0}(\beta):=\rho - \rho_{c}(\beta)> 0$ (Bose-Einstein condensation), whereas the rest of the sum converges to the
integral (\ref{rho<rho-c}) for $\mu = 0$, i.e. to $\rho_{c}(\beta)$. Therefore, by (\ref{Kac2}) and (\ref{rho>rho-c}) one
obtains the limit
\begin{eqnarray}\label{Kac4}
&&\lim_{V\rightarrow\infty}\mathbb{E}_{\beta, \, \mu_{L}(\beta, \rho)}(e^{- t \, N_{\Lambda}(\xi)/V})=
\int_{\rho_{c}(\beta)}^{\infty} dx \, \frac{e^{-(x - \rho_{c}(\beta))/(\rho - \rho_{c}(\beta))}}{\rho - \rho_{c}(\beta)}
\ e^{-t x}\\
&&= \frac{e^{-t \rho_{c}(\beta)}}{1 + t \, (\rho - \rho_{c}(\beta))}\ , \nonumber
\end{eqnarray}
where $\lim_{V\rightarrow\infty} \mu_{L}(\beta, \rho) =0$.

\begin{remark}\label{Kac-1} Representations (\ref{Kac3}), (\ref{Kac4}) and Definition \ref{mu-def} motivate the notion of
the Kac probability kernels $\{\mathcal{K}_{\beta, \, \overline{\mu}(\beta,\rho)}(x,\rho)\}_{\rho \geq 0}$, see \cite{LePu}:
\begin{eqnarray}\label{Kac5}
&&\mathcal{K}_{\beta, \, \overline{\mu}(\beta,\rho)}(x,\rho): = \theta_{+}(\rho_{c}(\beta) -\rho) \,
\delta_{\rho(\beta,\overline{\mu})}(x) + \\
&&\theta_{-}(\rho - \rho_{c}(\beta))\
\frac{\theta_{-}(x - \rho_{c}(\beta))\ e^{-(x - \rho_{c}(\beta))/(\rho - \rho_{c}(\beta))}}{\rho - \rho_{c}(\beta)} \ .
\nonumber
\end{eqnarray}
Here $\theta_{+}(z \geq 0) = 1$, $\theta_{+}(z < 0) = 0$ and $\theta_{-}(z > 0) = 1$, $\theta_{-}(z \leq 0) = 0$.
\end{remark}
Application of (\ref{Kac5}) to calculation of expectations in the \textit{grand-canonical} ensemble
\begin{equation}\label{Kac6}
\mathbb{E}_{\beta, \, \mu=\overline{\mu}(\beta,\rho)}(A)=
\int_{0}^{\infty} dx \, \mathcal{K}_{\beta, \, \overline{\mu}(\beta,\rho)}(x,\rho) \ \mathbb{E}_{\beta,\, x}(\hat{A}) \ ,
\end{equation}
shows that when $\overline{\mu}(\beta,\rho > \rho_{c}(\beta))= 0$, i.e. in the phase with \textit{condensate} density
$\rho_{0}(\beta)= \rho - \rho_{c}(\beta)$, the grand-canonical Gibbs state
\begin{equation}\label{Kac7}
\mathbb{E}_{\beta, \, \overline{\mu}(\beta,\rho> \rho_{c}(\beta))=0}(A)=
\int_{0}^{\infty} dx \, \mathcal{K}_{\beta, \, 0}(x,\rho) \ \mathbb{E}_{\beta,\, x}(\hat{A}) \ ,
\end{equation}
is \textit{convex} combination of the \textit{canonical} states $\{\mathbb{E}_{\beta, \, x}(\hat{A})\}_{x > \rho_{c}(\beta)}$.
Here the measure $dx \, \mathcal{K}_{\beta, \, 0}(x,\rho > \rho_{c}(\beta))$ defines the Kac probability distribution
with support $(\rho_{c}(\beta), +\infty)$, see (\ref{Kac5}).
\begin{remark}\label{Kac-2} Since the grand-canonical particle density $N_{\Lambda}(\xi)/V$ is non-negative random variable,
the Laplace transforms (\ref{Kac3}) and (\ref{Kac4}) defines characteristic function of the Kac distribution (\ref{Kac5}),
\cite{BLP}. Notice that $e^{- \alpha t \, \rho(\beta,\mu)}, \ \mu < 0$, and $\{1 + t \,
(\rho - \rho_{c}(\beta))\}^{-\alpha}, \  \rho \geq \rho_{c}(\beta)$ are characteristic functions for any $\alpha > 0$.
Therefore, two probability distribution with densities:
\begin{equation}\label{K-rho<rho-c}
\mathcal{K}_{\beta, \, \overline{\mu}(\beta,\rho)}(x,\rho) = \delta_{\rho(\beta,\overline{\mu})= \rho}(x) \ \theta_{+}(x)
\ , \ \rho \leq \rho_{c}(\beta)  \ ,
\end{equation}
see Definition \ref{mu-def},  and
\begin{equation}\label{K-BEC}
\mathcal{K}_{\rho_{c}(\beta)}^{BEC}(x,\rho):= \frac{\ e^{- x /(\rho - \rho_{c}(\beta))}}{\rho - \rho_{c}(\beta)} \
\theta_{-}(x)  \ , \ \rho > \rho_{c}(\beta) \ ,
\end{equation}
are infinitely divisible \cite{Sh}.
\end{remark}
Notice that (\ref{K-rho<rho-c}) coincides with the Kac probability kernels (\ref{Kac5})for $\rho \leq \rho_{c}(\beta)$. On
the other hand, by virtue of (\ref{Kac5}) and (\ref{K-BEC}) the Kac probability kernel for $\rho > \rho_{c}(\beta)$ is
\textit{convolution} of two probability measures:
\begin{eqnarray}\label{convol}
&&\mathcal{K}_{\beta, \, \overline{\mu}(\beta,\rho> \rho_{c}(\beta))}(x,\rho)=
(\mathcal{K}_{\rho_{c}(\beta)}^{BEC} \ast \mathcal{K}_{\beta, \, \overline{\mu}(\beta,\rho_{c})})(x,\rho) = \\
&&\int_{0}^{x} dy \ \mathcal{K}_{\rho_{c}(\beta)}^{BEC}(x-y ,\rho) \
\mathcal{K}_{\beta, \, \overline{\mu}(\beta,\rho_{c})}(y,\rho) = \int_{0}^{x} dy \
\mathcal{K}_{\rho_{c}(\beta)}^{BEC}(x-y ,\rho) \
\delta_{\rho_{c}(\beta)}(y) \ , \nonumber
\end{eqnarray}
see Definition \ref{mu-def}. Since the convolution of two infinitely divisible distribution is again infinitely divisible,
we deduce this property for the Kac distribution (\ref{Kac5}) also in the BEC phase with condensate density
$\rho_{0}(\beta)= \rho - \rho_{c}(\beta)>0$.\\
\textbf{(e) Spacial distribution of Boson particles.}
Since our aim is to analyse of the space distribution of boson for different initial geometry, we would like to indicate
how sensitive might be the Bose condensate shape as a function of the boundary conditions in the initial restricted geometry.
This implies certain specificity of position distribution of bosons even in the \textit{macroscopic} scale and in the
thermodynamic limit see e.g. \cite{VVZ}, \cite{Ver}.

Recall that for the \textit{periodic} boundary conditions the spectrum and the eigenfunctions are different to
(\ref{symm-spectr}) and (\ref{symm-eigenfunct}) for Dirichlet boundary conditions:
\begin{equation}\label{symm-spectr-per}
       \bigg\{ \varepsilon_\mathbf{k} = \frac{\hbar^2}{2m}\sum_{j=1}^3
          \Big( \frac{2\pi k_j}{L_j} \Big)^2 \, \bigg| \,
         \mathbf{k} = (k_1, k_2, k_3) \in \mathbb{Z}^3 \bigg\} \ ,
\end{equation}
\begin{equation}\label{symm-eigenfunct-per}
\bigg\{ \psi_{\mathbf{k}, \Lambda}(x) = \prod_{j=1}^3\sqrt{\frac{1}{L_j}}\
           \exp \Big(i \frac{2\pi k_j}{L_j}(x_j + L_j/2)\Big) \,
         \bigg| \, \mathbf{k} \in \mathbb{Z}^3 \bigg\} \ .
\end{equation}
Since the torus is wrapped,  the kernel of the corresponding operator $K_\Lambda$ is translation-invariant:
\begin{equation}\label{ODLROkernel-period}
K_{\Lambda}^{per}(x-y):= \frac{1}{L_1 L_2 L_3} \
\sum_{\textbf{k} \in \mathbb{Z}^3} \frac{\prod_{j=1}^3 \ e^{i \, {2\pi k_j}\ (x_j - y_j)/{L_j}}}
{\exp \beta [({\hbar^2 (2\pi)^2}/{2m})\sum_{j=1}^3 {k_j^2}/{L_j^2} +
\Delta]-1} \ ,
\end{equation}
where $\Delta =-\mu \geq 0$, since in this case $\varepsilon_{\mathbf{k}=(0,0,0)} =0$.
\begin{remark}\label{Per-D}
The kernels (\ref{ODLROkernel}) and (\ref{ODLROkernel-period}) produce different "global" position distributions of bosons
and the shape of the condensate. It is visible from ground state density
\begin{equation}\label{GrSt-D}
|\phi_{\mathbf{k}=\bf{1}, \Lambda}(x)|^2 = \frac{8}{L_1 L_2 L_3} \prod_{j=1}^3 \ \cos^2 (\pi x_j / L_j) \ ,
\end{equation}
for Dirichlet boundary conditions (\ref{symm-eigenfunct}) and
\begin{equation}\label{GrSt-Per}
|\psi_{\mathbf{k}=\bf{1}, \Lambda}(x)|^2 = \frac{1}{L_1 L_2 L_3} \ ,
\end{equation}
for the periodic boundary conditions. Hence one has to anticipate the difference in the local (in the vicinity of $x=0$)
as well as in the scaled space distributions. Below we essentially stick to the Dirichlet boundary conditions.
\end{remark}
\textbf{(f) BRPF theory of the conventional Bose-Einstein Condensation \cite{1,2}.} A peculiarity of the Tamura-Ito
approach \cite{1,2} is that BRPF is studied in the \textit{canonical} Gibbs ensemble (\ref{CE-2}). To take into account
the phenomenon of the Bose-Einstein condensation (BEC) they consider thermodynamic limit of the BRPF, when
$\Lambda \rightarrow \mathbb{R}^d$ for a fixed particle density $n/|\Lambda| = \rho$. The BRPF is completely
characterised by the limit of generating functional (\ref{CE-2}), and in \cite{1},\cite{2} this limit is established along
the family of $\Lambda\rightarrow \mathbb{R}^d$, which are \textit{non-anisotropic}. This is a standard precaution to
reach the conventional bulk properties and in particular the \textit{conventional} BEC. It is known as the
\textit{Fisher limit}, for sufficient conditions see e.g. \cite{Ru}.

The first result \cite{1} concerns the case of \textit{subcritical} density of bosons $\rho < \rho_{c}(\beta)$, cf.(d).
Then for any non-negative continuous function $f$ with a compact support the finite volume \textit{canonical} BRPFs
$\{\mu_{\Lambda, n}\}_{\Lambda}$ converge \textit{weakly} to the random point field $\mu_{\rho}$ such that the limit
generating functional (\ref{CE-2}) takes the form:
\begin{equation}\label{CE-rho<rho-c}
\mathcal{Z}_{\rho}[f] := \lim_{\Lambda \rightarrow \mathbb{R}^d : \, \rho = n/V}\mathcal{Z}_{\Lambda, n}[f] =
\int_{Q(\mathbb{R}^d)} \mu_{\rho}(d\xi) \, e^{-\langle f, \xi \rangle}
= \Det[1 + K_{f}({z(\rho)})]^{-1} \ \ .
\end{equation}
Here $K_{f}({z}):= \sqrt{1-e^{-f}} K(z) \sqrt{1-e^{-f}}$ and $K(z):= \, {z} G_\beta (1- {z} G_\beta)^{-1} $, where
$G_\beta:= e^{\beta \Delta}$ is the heat semigroup in the whole space $\mathbb{R}^d$, cf. (\ref{Kf-z}), and $z(\rho)$
is a unique solution of equation
\begin{equation}\label{z(rho)}
\rho = \frac{1}{(2 \pi)^3}\int_{\mathbb{R}^3}\frac{d^3 k}{z^{-1} \, e^{\beta\epsilon_{\textbf{k}}}-1} \ ,
\end{equation}
for the total particle density $\rho \leq \rho_{c}(\beta)$. Similar to the grand-canonical ensemble, see
(\ref{rho<rho-c}), (\ref{rho-crit}), $\lim_{\rho\rightarrow\rho_{c}(\beta)} z(\rho)= 1$ and equation (\ref{z(rho)})
has no solution for $\rho > \rho_{c}(\beta)$. The excess of the particle density $\rho_{0}(\beta):=\rho - \rho_{c}(\beta)> 0$
accumulates in the ground state ${\bf{1}} = (1,1,1)$ (Bose-Einstein condensation), although absence of the explicit
formula (\ref{meanPartDens}) makes this analysis more complicated than in grand-canonical approach \cite{PuZ}.

In spite of these difficulties Tamura and Ito succeeded to find the limit of the \textit{canonical} BRPFs
$\{\mu_{\Lambda, n}\}_{\Lambda}$ in the phase with condensation \cite{2}:
\begin{eqnarray}\label{CE-rho>rho-c}
&&\mathcal{Z}_{\rho}[f] := \lim_{\Lambda \rightarrow \mathbb{R}^d : \, \rho = n/V > \rho_{c}}\mathcal{Z}_{\Lambda, n}[f] =
\int_{Q(\mathbb{R}^d)} \mu_{\rho}^{c}(d\xi) \, e^{-\langle f, \xi \rangle} = \\
&&\frac{\exp \{-(\rho - \rho_{c}(\beta))(\sqrt{1-e^{-f}}, [1+ K_{f}(1)]^{-1} \sqrt{1-e^{-f}})\}}{\Det[1 + K_{f}(1)]}
\nonumber \ .
\end{eqnarray}
Their detailed analysis of the canonical limit shows that the structure of generating functional (\ref{CE-rho>rho-c})
is due to \textit{convolution}: $\mu_{\rho}^{c} = \mu_{\rho_{c}(\beta)}\ast \mu_{\rho-\rho_{c}(\beta)}^{0}$ of
\textit{two} boson random point processes. The first one is (\ref{CE-rho<rho-c}) for the critical density
$\rho= \rho_{c}(\beta)$, whereas the second one $\mu_{\rho_{0}(\beta)}^{0}$ corresponds to the BRPF of particles in the
condensate with density $\rho_{0}(\beta):=\rho - \rho_{c}(\beta)> 0$ :
\begin{equation}\label{CE-c}
\int_{Q(\mathbb{R}^d)}\mu_{r}^{0}(d\xi)\ e^{-\langle f, \xi \rangle}
= \exp\{-{r} (\sqrt{1-e^{-f}}, \frac{1}{1+K_{f}({1})}\sqrt{1-e^{-f}}) \} \ , \ r>0.
\end{equation}

Finally, taking into account (\ref{gf0}) and definitions (\ref{GC-PartFunc}),(\ref{GC-PartFunc-f}) together with
(\ref{CanonPart}), (\ref{CanonPart-f}), (\ref{CE-2}), we obtain representation
\begin{equation}\label{Kac2-f}
\mathcal{Z}_{\Lambda, \mu}^{gc}[f]=
\sum_{n=0}^{\infty}  \ \frac{e^{\beta n \mu }\ {Z}_{\Lambda}(\beta, n)}{\Xi_{\Lambda}(\beta, \mu)} \
\mathcal{Z}_{\Lambda, n}[f]  \ .
\end{equation}
Notice that discrete measures in (\ref{Kac2}) and (\ref{Kac2-f}) are identical. Therefore, in the thermodynamic limit
the grand-canonical
generating functional $\mathcal{Z}_{\mu=\overline{\mu}(\beta,\rho)}^{gc}[f]$ is related to the canonical one by
the Kac transformation:
\begin{equation}\label{Kac3-f}
\mathcal{Z}_{\mu=\overline{\mu}(\beta,\rho)}^{gc}[f] =  \int_{0}^{\infty} dx \, \mathcal{K}_{\beta, \,
\overline{\mu}(\beta,\rho)}(x,\rho)\
\mathcal{Z}_{x}[f] \ .
\end{equation}
Then taking into account (\ref{Kac5}) this implies
\begin{eqnarray}\label{Kac4-f}
&&\mathcal{Z}_{\mu=\overline{\mu}(\beta,\rho)}^{gc}[f] = \theta_{+}(\rho_{c}(\beta) -\rho) \, \Det[1 + K_{f}({z(\rho)})]^{-1}
+ \\
&&\theta_{-}(\rho - \rho_{c}(\beta))\ \frac{\Det[1+K_{f}(1)]^{-1}}
{1+ (\rho - \rho_{c}(\beta))(\sqrt{1-e^{-f}},(1+K_{f}(1))^{-1} \sqrt{1-e^{-f}})} \ . \nonumber
\end{eqnarray}
Generating functional (\ref{Kac4-f}) expresses  the Tamura-Ito results \cite{1,2} in the grand-canonical ensemble.

\subsection{Microscopic and macroscopic distributions of constituent bosons
in exponentially anisotropic SLAB}\label{sub-sec-SLAB}
\textbf{(a) Two critical points and generalised BEC in anisotropic SLAB.}
First, we describe the space distribution of the constituent bosons in the thermodynamic limit of \textit{exponentially}
anisotropic (in $x_1 , x_2$-directions) SLAB with Dirichlet boundary conditions.

To this end we set $L_1 = L_2 = L \, e^{\alpha L}, L_3 = L$ for $\alpha >0$ and consider the limit $L \rightarrow \infty$.
We also denote this parallelepiped by $\Lambda =\Lambda_L$ and by $V_L = L^3 \, e^{\alpha L}$ its volume.

Let $\mu\leq 0$. Then for the limits of the Darboux-Riemann sums (\ref{meanPartDens}) one gets:
\begin{eqnarray}
&&\lim_{L\rightarrow\infty}\frac{1}{V_L} \sum_{\textbf{k}\in \mathbb{N}^3}
\frac{1}{\exp \beta [({\hbar^2\pi^2}/{2m})\sum_{j=1}^3 {(k_j^2-1)}/{L_j^2} +
\Delta ]-1} = \nonumber \\
&&\lim_{L\rightarrow\infty}\frac{1}{V_L} \sum_{\textbf{k} \neq (k_1,k_2,1)}
\frac{1}{\exp \beta [({\hbar^2\pi^2}/{2m})\sum_{j=1}^3 {(k_j^2-1)}/{L_j^2} +
\Delta ]-1} = \nonumber \\
&&\frac{1}{(2 \pi)^3}\int_{\mathbb{R}^3}\frac{d^3 k}{e^{\beta(\hbar^2 k^2/2m + \Delta)}-1} =: \rho(\beta,\mu = -\Delta)\ .
\label{DR-n1n2}
\end{eqnarray}
Here $\rho(\beta,\mu)$ is the the {grand-canonical} particle density mean-value with finite critical density
$\rho_c (\beta):= \rho(\beta,\mu=0)$. Since (\ref{DR-n1n2}) yields:
\begin{equation}\label{Eq1}
\lim_{L\rightarrow\infty}\sum_{\textbf{k} \neq (k_1,k_2,1)} \frac{1}{\exp \beta [({\hbar^2\pi^2}/{2m})
\sum_{j=1}^3 {(k_j^2-1)}/{L_j^2}]-1} = \rho_c (\beta) \ ,
\end{equation}
by virtue of (\ref{meanPartDens}) for $\rho > \rho_c (\beta)$ we obtain that:
\begin{eqnarray}
&&\rho - \rho_c (\beta) = \lim_{L\rightarrow\infty} \frac{1}{V_L} \sum_{\textbf{k}=(k_1,k_2,1)}
\frac{1}{\exp \beta [({\hbar^2\pi^2}/{2m})\sum_{j=1}^3 {(k_j^2-1)}/{L_j^2} + \Delta_L ]-1} =  \nonumber \\
&&\lim_{L\rightarrow\infty} \frac{1}{L} \frac{1}{(2 \pi)^2}\int_{\mathbb{R}^2}\frac{d^2k}
{e^{\beta(\hbar^2 k^2/2m + \Delta_{L}(\beta,\rho))}-1} =
\lim_{L\rightarrow\infty} \frac{1}{\lambda_{\beta}^{2} L} \ln [\beta \Delta_{L}(\beta,\rho)]^{-1} \ .\label{Eq2}
\end{eqnarray}
This implies the asymptotics:
\begin{equation}\label{ro<roM}
\Delta_{L}(\beta,\rho) = \frac{1}{\beta} \  e^{- \lambda_{\beta}^{2} (\rho - \rho_c (\beta)) L } + o(e^{-A L}) \ .
\end{equation}
Notice that representation of the limit (\ref{Eq2}) by the integral is valid only when
$\lambda_{\beta}^{2}(\rho - \rho_c (\beta)) < 2 \alpha$. For $\rho$ larger than the
\textit{second} critical density \cite{BZ}:
\begin{equation}\label{roM}
\rho_m(\beta):= \rho_c (\beta) + 2 \alpha / \lambda_{\beta}^{2}
\end{equation}
the correction $\Delta_{L}(\beta,\rho)$ must converge to zero faster than $e^{- 2\alpha L }$. Now to keep the
difference positive: $\rho - \rho_m (\beta) > 0$, we have to return back to the original sum representation
(\ref{Eq1}) and (as for the standard BEC) to take into account the impact of the ground state occupation density
\textit{together} with a saturated \textit{non-ground state} (i.e. generalised) condensation $\rho_m(\beta)- \rho_c (\beta)$
as in (\ref{Eq2}). For this case the asymptotics of $\Delta_{L}(\beta,\rho > \rho_m(\beta))$ is
completely \textit{different} than (\ref{ro<roM}) and it is equal to
\begin{equation}\label{ro>roM}
\Delta_{L}(\beta,\rho) = [\beta (\rho - \rho_m(\beta))V_L]^{-1} + o(V_{L}^{-1})\ .
\end{equation}
Since $V_L= L^3 \, e^{2\alpha L}$, we obtain:
\begin{eqnarray}
&&\lim_{L\rightarrow\infty} \frac{1}{V_L} \sum_{\textbf{k} =(k_1 > 1,k_2 > 1, 1)}
\frac{1}{\exp \beta [({\hbar^2\pi^2}/{2m})\sum_{j=1}^3 {(k_j^2-1)}/{L_j^2} + \Delta_L ]-1}= \nonumber \\
&&\lim_{L\rightarrow\infty} \frac{1}{\lambda_{\beta}^{2} L} \ln [\beta \Delta_{L}(\beta,\rho)]^{-1} =
2 \alpha / \lambda_{\beta}^{2} = \rho_m(\beta)- \rho_c (\beta), \label{Eq3}
\end{eqnarray}
and there is a \textit{macroscopic} occupation of the ground-state that yields for the term corresponding to
$\textbf{k} =(1,1,1)$:
\begin{equation}\label{ro>roM}
\rho - \rho_m (\beta)= \lim_{L\rightarrow\infty}
\frac{1}{V_L}\frac{1}{e^{\beta \Delta_{L}(\beta,\rho)}-1} > 0 \ .
\end{equation}

Notice that for $\rho_c(\beta)< \rho< \rho_m (\beta)$ we obtain the vdBLP-gBEC
(of the \textit{type} III), i.e. \textit{none} of the single-particle states are \textit{macroscopically}
occupied, since the exponential anisotropy: $L_1 = L_2 = L \, e^{\alpha L}, L_3 = L$,  and (\ref{ro<roM}) imply for any
mode $\textbf{k}$:
\begin{equation}\label{typeIII}
\rho_{\textbf{k}} (\beta,\rho):= \lim_{L\rightarrow\infty}
\frac{1}{V_L}\frac{1}{e^{\beta(\epsilon_{\textbf{k}} - \mu_{L}(\beta,\rho))}-1} = 0 \ .
\end{equation}
On the other hand, the asymptotics $\Delta_{L}(\beta,\rho> \rho_m(\beta)) =
[\beta (\rho - \rho_m(\beta))V_L]^{-1}$
implies
\begin{equation}\label{typeI-typeIII}
\rho_{\textbf{k}\neq (1,1,1)} (\beta,\rho):= \lim_{L\rightarrow\infty}
\frac{1}{V_L}\frac{1}{e^{\beta(\epsilon_{\textbf{k}} - \mu_{L}(\beta,\rho))}-1} = 0 \ ,
\end{equation}
i.e. for $\rho > \rho_m(\beta)$ there is a \textit{coexistence} of the \textit{saturated} type
III gBEC, with the \textit{constant} density (\ref{Eq3}), and the standard, i.e. the \textit{type} I vdBLP generalised
BEC in the single (ground) state (\ref{ro>roM}), \cite{BZ}, \cite{MuSa}.

Since the inverse temperature $\beta$ and the chemical potential $\mu$ are the intrinsic parameters of the Gibbs
grand-canonical ensemble, it is natural to use them in our analysis of BRPF in the thermodynamic limit. However, as
it is indicated above to ketch the BEC in this ensemble we have to make the chemical potential $L$-dependent.
To elucidate this $L$-dependence for anisotropic boson systems we consider first the average density
$\bar\rho( L, \Delta)$, i.e., the  Gibbs expectation value of the total particle number in reservoir $\Lambda_L$
divided by the volume $V_L$:
\begin{equation}
\bar\rho( L, \Delta) := \frac{1}{V_L}
\sum_{\textbf{k}\in \mathbb{N}^3} W(k, L, \Delta) \ ,
\label{DR-n1n2}
\end{equation}
where
\begin{equation}
W(k, L, \Delta) := \frac{1}{\exp [({\beta \hbar^2\pi^2}/{2m})\sum_{j=1}^3
{(k_j^2-1)}/{L_j^2} + \beta \Delta]-1} .
\end{equation}
Here and hereafter, we set $\beta > 0$ fixed, but arbitrary,
and suppress it as argument if it will not produce any confusion.

Recall that $\mu = \epsilon_{\bf 1}(L) - \Delta$ and the condition
$\Delta >0$ is needed for definiteness (\textit{stability}) of the ideal Bose-gas.
Then for fixed $\Delta >0$ ( or for fixed $\mu < \epsilon_{\bf 1}(L)$), we obtain
in the limit
\[
\bar\rho(\Delta) := \lim_{L\rightarrow\infty}
\bar\rho(L, \Delta) = \frac{1}{(2 \pi)^3}\int_{\mathbb{R}^3}
\frac{dk}{e^{\beta(\hbar^2 k^2/2m + \Delta)}-1}
< \frac{1}{(2 \pi)^3}\int_{\mathbb{R}^3}
\frac{dk}{e^{\beta\hbar^2 k^2/2m}-1} = \bar\rho_c
\]
and $\lim_{\Delta\downarrow 0}\bar\rho(\Delta) = \bar\rho_c$.

To make $\bar\rho(\Delta) > \bar\rho_c$, we let $\Delta$ be
$L$-dependent.
In fact, if $\lim_{L\to\infty}\Delta(L) = 0$, we have
\[
      \lim_{L\to\infty} \frac{1}{V_L}
          \sum_{\textbf{k}\in \mathbb{N}^3, k_3>1 }
         W(k, L, \Delta(L)) = \bar\rho_c
\]
and
\[
      \lim_{L\to\infty} \frac{1}{V_L}
          \sum_{\textbf{k}\in \mathbb{N}^3, k_3=1 }
         W(k, L, \Delta(L)) =
          \lim_{L\to\infty} \frac{1}{V_L}
          \Big\{ W({\bf 1}, L, \Delta(L)) +
       \sum_{\textbf{k}\in \mathbb{N}^3,(k_1,k_2) \ne(1,1), k_3=1 }
         W(k, L, \Delta(L))\Big\}
\]
\begin{equation}
      = \lim_{L\to\infty}\Big\{ \frac{1}{L^3e^{2\alpha L}\beta\Delta(L)}
         +\frac{m}{2\beta \hbar^2\pi L}
       \log(L^2e^{2\alpha L}\wedge \Delta(L)^{-1}) + O(L^{-1})\Big\},
\label{avedsty}
\end{equation}
see (\ref{A1}).
If we set $\Delta(L) =e^{-\gamma L} \; (0 < \gamma \leqslant 2\alpha)$,
for example, we have
\[
    \bar\rho(\Delta) =  \lim_{L\to\infty}\bar\rho(L,\Delta(L))
       = \frac{m\gamma}{2\beta\hbar^2\pi} + \bar\rho_c \leqslant
      \frac{2m\alpha}{2\beta\hbar^2\pi} + \bar\rho_c =:\bar\rho_m.
\]
In this $L$-dependence on $\Delta$,
we obtain the type-III gBEC, i.e. \textit{none}
of the single-particle states are \textit{macroscopically} occupied,
and there are macroscopic contributions to
the averaged density from infinitely many but relatively small parts of
single-particle states whose quantum numbers lay in $(\N^2 -\{(1,1)\})
\times \{1\}$ out of $\N^3$.
Note that the first term of (\ref{avedsty}) is zero and the second term
yields $\bar\rho(\Delta) - \bar\rho_c$.

If we set $\Delta(L) = (L^3e^{2\alpha L}\delta)^{-1} \quad (\delta >0)$,
we get
\[
      \lim_{L\to\infty}\bar\rho(L,\Delta(L)) = \bar\rho_m + \delta,
\]
which indicates that there is a \textit{coexistence} of the
\textit{saturated} type III vdBLP generalised BEC and the standard
(i.e. the type I vdBLP in the single-particle ground state, \cite{BZ}, \cite{MuSa}).
Note that the first term of (\ref{avedsty}) yields
$\bar\rho(\Delta) - \bar\rho_m$ and the second term yields
$\bar\rho_m - \bar\rho_c$.

\noindent \textbf{(b) Spacial distribution of BRPF in anisotropic SLAB.}
Now let us examine the detailed position distribution of these \textit{three} phases of the exponential SLAB system.
The above observation indicates that it is convenient to introduce the parameters:
\begin{equation}\label{kappaf}
\kappa_1 := \lim_{L\to\infty} \frac{8}{\beta\Delta(L) L^3e^{2\alpha L}} , \qquad
\kappa_2 := \lim_{L\to\infty}
\frac{m}{\beta\pi\hbar^2L}\log\big[(L^2e^{2\alpha L})
\wedge (\beta\Delta(L))^{-1}\big]
\end{equation}
to distinguish these phases. We restrict our attention to those asymptotics for $\Delta(L)$, that give finite
values of $ \kappa_1$ and $ \kappa_2$.
Note that there are differences in factors in $\kappa_1$ and
$\kappa_2$ with those quantities in (\ref{avedsty}).
We use the present form to make the results in the following theorems
simple.
The differences in the factors occur because we consider the different
object in the following argument from the above.
It is the thermodynamic limit of the local density near the coordinate origin rather
than the density obtained by averaging over the region $\Lambda_L$.
Correspondingly, there is a difference between critical values
$\bar\rho_c, \bar\rho_m$ of the averaged density and
$\rho_c, \rho_m$ of our BRPF discussed below.
In fact, we have $\rho_m \ne \bar\rho_m$, while $\rho_c = \bar\rho_c$.
See Remark \ref{rho} and Remark \ref{density2}.

First, we consider the \textit{microscopic} distribution of the constituent bosons in terms of BRPF.
The result is the following theorem:
\begin{thm}
If $\lim_{L\to \infty} \Delta(L) =\Delta_{\infty} >0 $,
then the BRPFs $\{ \, \nu_{\mu,\Lambda_L}\, \}_{L>0} $ converge weakly
to the BRPF $\nu_{\Delta_{\infty}, 0}$.
If $ \Delta_{\infty} = 0 $, then the BRPFs $\{ \, \nu_{\mu,\Lambda_L}\, \}_{L>0} $ converge weakly
to the BRPF $\nu_{0, \kappa}$ for $  \kappa = \kappa_1 + \kappa_2. $
Here $\nu_{\Delta_{\infty}, 0}$ is the BRPF characterised by the
generating functional:
\[
          \int_{Q(\R^3)}e^{-\langle f, \xi \rangle}
        \nu_{\Delta_{\infty}, 0}(d\xi) =
      \Det[1+K_f^{\Delta_{\infty}}]^{-1} \ ,
\]
with $K_f^{\Delta_{\infty}}= \sqrt{1-e^{-f}}\, G(e^{\beta\Delta_{\infty}}-G)^{-1}\sqrt{1-e^{-f}}$.
Whereas the BRPF $\nu_{0, \kappa}$ is characterised by the generating functional:
\[
          \int_{Q(\R^3)}e^{-\langle f, \xi \rangle}
        \nu_{0, \kappa}(d\xi) = \big(1+\kappa(\sqrt{1-e^{-f}},
     (1+K_f^0)^{-1} \sqrt{1-e^{-f}} \, )\big)^{-1}
      \Det[1+K_f^0]^{-1},
\]
with $K_f^0 = \sqrt{1-e^{-f}}\, G(1-G)^{-1} \sqrt{1-e^{-f}}= K_f^{\Delta_{\infty}=0} $, where $f \in C_0(\R^3)$ is
non-negative and  $G=e^{\beta\hbar^2\triangle/2m}$ is the heat semigroup on the whole space $\R^3$.
\label{thmA}
\end{thm}
\begin{remark}
In the case $\Delta_{\infty}=0$, the operator
$G(e^{\beta\Delta_{\infty}}-G)^{-1}$ is unbounded.
For a careful argument of this case, we refer to \cite{2}.
\end{remark}
\begin{remark}
Let $\mu_n, \mu_{\kappa}, \mu_r^{(ex)}$ be BRPFs whose generating functionals
are given by
\[
         \int_{Q(\R^3)}e^{-\langle f, \xi \rangle}
        \mu_{n}(d\xi) = \Det[1+K_f^0]^{-1},
\]
\[
          \int_{Q(\R^3)}e^{-\langle f, \xi \rangle}
        \mu_{\kappa}(d\xi) = \big(1+\kappa(\sqrt{1-e^{-f}}, \,
     (1+K_f^0)^{-1} \sqrt{1-e^{-f}} \, )\big)^{-1}
\]
and
\[
       \int_{Q(\R^3)}e^{-\langle f, \xi \rangle}
        \mu_r^{(ex)}(d\xi) = e^{-r(\sqrt{1-e^{-f}}, \,
     (1+K_f^0)^{-1} \sqrt{1-e^{-f}} \, )} .
\]
Then we have the convolution
\[
        \nu_{0, \kappa} = \mu_n * \mu_{\kappa}
\]
and the decomposition into ergodic components
\[
        \mu_{\kappa} = \int_0^{\infty}
       e^{-t}\mu^{(ex)}_{\kappa t}dt,
\]
which shows that $\nu_{\kappa}$ is a mixture of pure states
$\mu_n*\mu_{\kappa t}^{(ex)} \; (t>0)$.
{This mixture property would be grand canonical ensemble artifact.
Then this is a manifestation of the Kac decomposition.}
\end{remark}
\begin{remark}
{From} the above expression of the generating functional, we have
\[
             \int_{Q(\R^3)} \langle f, \xi \rangle \, \nu_{0,\kappa}(d\xi)
        = \int_{\R^3} \rho f(x) \, dx
\]
for $\rho = \kappa_1 + \kappa_2 + K^{\Delta_{\infty}}(x, x)$,
where $K^{\Delta_{\infty}}(x, y)$ is the kernel of the operator
$G(e^{\beta\Delta_{\infty}}-G)^{-1}$, i.e.,
\begin{equation}
      K^{\Delta_{\infty}}(x, y) = \int_{\R^3}\frac{e^{ik\cdot (x-y)}}
     {e^{\beta(\hbar^2|k|^2/{2m}+\Delta_{\infty})} -1}
       \, \frac{dk}{(2\pi)^3}.
\label{Kxy}
\end{equation}
Note that $K^{\Delta_{\infty}}(x, x) $ is $x$-independent.
Hence, $\rho$ may be regarded as the density of the system, which is
\textit{uniform}. This uniformity is the result of the thermodynamic limit.

In fact, $\rho$ corresponds to the local density of the compact region
near the origin before thermodynamic limit, where the local density of the condensed
part is expected to depend on the position because the ground state
wave function of the Dirichlet boundary conditions is not uniform.
When $\Delta_{\infty} > 0$, $ \kappa_1 = \kappa_2 = 0 $ and
$ \rho < \rho_c $.
While $\rho \geqslant \rho_c$ corresponds to the $\Delta_{\infty}=0$
case, where
\[
          \rho_c = \bar\rho_c = \int_{\R^3}\frac{1}
     {e^{\beta\hbar^2|k|^2/{2m}} -1}
       \, \frac{dk}{(2\pi)^3}
\]
\label{rho}
\end{remark}
\begin{remark}
As one can see from the argument above and the proof of the theorem
that $\kappa_1$ is a contribution of the ground state
$\phi_{{\bf 1}, \Lambda}$, $\kappa_2$
comes from the Riemann sum which is the sum of contributions of the
states $\{\phi_{ \, (s_1,s_2,1), \Lambda} \, | \, s_1, s_2 \in \N,
(s_1, s_2) \ne (1, 1) \,\}$ and $K(x,x)$ is a contribution of other states.
That is, if $\kappa_1 \ne 0$, the system enjoys a usual BEC
(i.e., type I of gBEC).
If $\kappa_2 \ne 0$, it is in type III gBEC. \cite{BLP}

\noindent This feature also holds for Theorem \ref{thmB} and
Theorem \ref{thmC}.
\end{remark}
\begin{remark}
In the above argument, we have derived the ``density" $\rho$ from the
view point that the system is specified by $\beta$ and the
$L$-dependence of $\Delta$ (i.e., $\mu$).
However it may be natural to understand that the $L$-dependence of
$\Delta$ is induced from $\beta$ and $\rho$.
{From} this point of view, for $(\beta, \rho)$ in the region
$\rho \leqslant \rho_c$, we may take $L$-independent
$\Delta = \Delta_{\infty}$
which is the unique solution of
\[
      \rho = \int_{\R^3}\frac{1}
     {e^{\beta(\hbar^2|k|^2/2m+\Delta_{\infty})} -1}
       \, \frac{dk}{(2\pi)^3};
\]
for $(\beta, \rho)$ in the region
$\rho_c < \rho \leqslant \rho_{\rm m}$,
$\Delta(\rho,\ L)$ is given by
\[
       \Delta(\rho, L) = \frac{1}{\beta}
 e^{-\beta\pi\hbar^2(\rho-\rho_c(\beta))L(1+o(1))/m},
\]
where $\rho_{\rm m} = 2m\alpha/(\beta\pi\hbar^2)$;

\noindent for $(\beta, \rho)$ in the region
$ \rho > \rho_{\rm m}$, $\Delta(\rho, L)$ is
given by
\[
      \Delta(\rho, L) = \frac{8}{\beta L^2e^{2\alpha L}
          (\rho - \rho_{\rm m}(\beta))}(1+o(1)).
\]
\end{remark}
\smallskip

Now let us consider the type of \textit{macroscopic} distribution of the
constituent particles in the exponential SLAB by using \textit{two different}
scaling arguments.

First, we deal with the scaling, in which $\Lambda_L$ can be seen in the thermodynamic limit as a two-dimensional
\textit{square} $S=[-1/2, 1/2]^2$, i.e. a \textit{finite} infinitely thin \textit{slab}. To this end we scale point
measures by the transformation
\[
     T_S: Q(\Lambda_L) \ni \xi = \sum_j \delta_{x^{(j)}} \longmapsto
             \eta = \frac{1}{L^3e^{2\alpha L}} \sum_j
      \delta_{(x_1^{(j)}/Le^{\alpha L}, \, x_2^{(j)}/Le^{\alpha L} )}
           \in {\cal M}(S) \ ,
\]
where ${\cal M}(S)$ is the space of all (locally) finite non-negative Borel measures on $S$.

Now let us introduce the following \textit{Random Field} (RF), i.e., a Borel
probability measure on ${\cal M}(S)$ :
\begin{equation}
\mathscr{M}_{\mu, L}^{(S)} = \nu_{\mu, \Lambda_L} \circ T_S^{-1} \ ,
\label{T}
\end{equation}
and also define the RF $\mathscr{M}_{a, b}^{(S)}$, which is characterised by the
generating functional
\[
          \int_{{\cal M}(S)}e^{-\langle f, \eta \rangle}
          \mathscr{M}_{a, b}^{(S)}(d\eta)
       =  \exp\Big[-a \int_{S} f(x) \, dx \Big]
\]
\[
       \times \Big[ 1+ b \int_{S}f(x)
         \big(\prod_{j=1}^2\cos^2\pi x_j \big) \, dx \Big]^{-1},
\]
for non-negative $f\in C(S)$. Then we get the following result:
\begin{thm}
The random measure $ \mathscr{M}_{\mu, L}^{(S)} $ converges to
$ \mathscr{M}_{a, b}^{(S)} $
with $  a = K^{\Delta_{\infty}}(x, x) +  \kappa_2/2 $
and  $  b =  \kappa_1/2  $ weakly, with
$K^{\Delta_{\infty}}(x,x)$ defined by (\ref{kappaf}) and (\ref{Kxy}).
\label{thmB}\end{thm}
\begin{remark}
Let $\delta_g$ be the ``deterministic" RF which has point mass
on the singleton $g\, dx \in {\cal M}(S)$, i.e.,
\[
       \int_{{\cal M}(S)} e^{-\langle f, \eta \rangle}
     \delta_g(d\eta) = e^{-\int_{S}f(x)g(x)\, dx}.
\]
Then, $\mathscr{M}_{a, b}^{(S)}$ has the decomposition to deterministic
RFs:
\[
      \mathscr{M}_{a, b}^{(S)} = \int_0^{\infty} \delta_{g_t}e^{-t}dt,
\]
where
\[
        g_t(x) = a + b \, t \big(\prod_{j=1}^2\cos^2\pi x_j\big).
\]
The first term of $g_t$ is a contribution of the mixture of the normal and the type-III gBEC
phases, while the second comes from the usual BEC, see Sec.\ref{sub-sec-SLAB},
Sec.\ref{sub-sec-BEAM} and in particular (\ref{kappaf}).

The $x_1, x_2$ dependence of the second term reflects the square of
the ground state wave function of the Dirichlet boundary conditions.
In fact, it is the scaled and squared ground state wave-function
after integrating the third variable $x_3$.
\label{remB1}
\end{remark}
\begin{remark}
We have
\[
      \int_{{\cal M}(S)}\langle f, \eta \rangle
          \mathscr{M}_{a, b}^{(S)}(d\eta)
       =   \int_{S} \rho(x) f(x) \, dx ,
\]
where
\[
         \rho(x) = K^{\Delta_{\infty}}(x, x) + \frac{\kappa_2}{2} +
        \frac{\kappa_1}{2}\prod_{j=1}^2\cos^2\pi x_j
\]
can be regarded as the density function of the system from the
macroscopic point of view.
Note that by averaging $\rho$, we have
\[
      \int_S \rho(x)\, dx/|S| =
   K^{\Delta_{\infty}}(x,x) + \frac{\kappa_2}{2} + \frac{\kappa_1}{8} = \bar\rho,
\]
where $|S| = 1$ is the Lebesgue measure of $S$.
This, the averaging density of grand canonical ensemble (\ref{avedsty})
is re-derived.
\label{density2}
\end{remark}

The second scaling we consider here concerns another {macroscopic} (\textit{mesoscopic})
distribution, where the system looks in the limit as an \textit{infinite} plane \textit{slab} with a \textit{finite}
thickness:
$D = {\Bbb R}^2\times [-1/2, 1/2]$.

Let us scale point measures by the following transformation:
\begin{equation}\label{T_D}
T_D: Q(\Lambda_L) \ni \xi = \sum_j \delta_{x^{(j)}} \longmapsto \eta =
\frac{1}{L^3} \sum_j \delta_{x^{(j)}/L } \in {\cal M}(D_L) \subset {\cal M}(D) ,
\end{equation}
where $D_L =[-e^{\alpha L}/2, e^{\alpha L}/2]^2\times [-1/2, 1/2]$ and
${\cal M}(D)$, or ${\cal M}(D_L)$, are the space of all locally finite non-negative
measures on $D$, or $D_L$, respectively.

Let us introduce a RF on ${\cal M}(D)$ by
\begin{equation}
        {\mathscr{M}}_{\mu, L}^{(D)} = \nu_{\mu, \Lambda_L} \circ T_D^{-1}.
\label{TD}
\end{equation}
We also define a RF $ {\mathscr{M}}_{a, b}^{(D)}$ on ${\cal M}(D)$  which characterized
by the generating functional
\[
          \int_{{\cal M}(D)}e^{-\langle f, \eta \rangle}
          {\mathscr{M}}_{a, b}^{(D)}(d\eta)
       =  \exp\Big[-a \int_{D} f(x) \, dx \Big]
\]
\[
       \times \Big[ 1+ b \int_{D}f(x)
         \cos^2\pi x_3 \, dx \Big]^{-1},
\]
where $f\in C_0(D)$ is non-negative.
We get the following result:
\begin{thm}
The random measure $ {\mathscr{M}}_{\mu, L}^{(D)} $ converges to
$ {\mathscr{M}}_{a, b}^{(D)} $
with
\[
      a = K^{\Delta_{\infty}}(x, x) \quad \mbox{ and } \quad
     b =  \kappa_1 + \kappa_2,
\]
weakly, with
$K(x,x)$ defined by (\ref{Kxy}).
\label{thmC}\end{thm}

\begin{remark}
Here $\kappa$ is same with that in Theorem \ref{thmA}.
We can also get a similar decomposition of the RF as in Remark \ref{remB1}.
\end{remark}

\begin{remark}
The ``density" function of this scale is
\[
          \rho(x) = K^{\Delta_{\infty}}(x,x) +
           (\kappa_1 + \kappa_2)\cos^2\pi x_3.
\]
The second term is proportional to the value at $(0, 0, x_3)$   of
scaled squared ground state wave function.
\end{remark}

\begin{remark}
It would help understanding the whole picture of the SLAB system
to summarize the results of the three scales roughly as
\begin{eqnarray}
          \int_{Q(\R^3)}e^{-\langle f, \xi \rangle}
        \nu(d\xi) &=& \big(1+(\rho - \rho_c)^+
        (\sqrt{1-e^{-f}},
     (1+K_f^{\Delta_{\infty}})^{-1} \sqrt{1-e^{-f}} \, )\big)^{-1}
\notag
\\
     & & \; \times \Det[1+K^{\Delta_{\infty}}_f]^{-1},
\label{mic}
\\
          \int_{{\cal M}(S)}e^{-\langle f, \eta \rangle}
          \mathscr{M}^{(S)}(d\eta)
       &=&  \exp\Big[-\frac{(\rho\wedge\rho_{\rm m} -
         \rho_c)^+}{2}
        + \rho\wedge\rho_c  \big) \int_{S} f(x) \, dx \Big]
\notag
\\
     & & \; \times \Big[ 1+ \frac{(\rho-\rho_{\rm m})^+}{2}
           \int_{S}f(x)
         \big(\prod_{j=1}^2\cos^2\pi x_j \big) \, dx \Big]^{-1},
\label{mac}
\\
          \int_{{\cal M}(D)}e^{-\langle f, \eta \rangle}
          {\mathscr{M}}^{(D)}(d\eta)
       &=&  \exp\Big[- (\rho\wedge\rho_c)\int_{D} f(x)
                 \, dx \Big]
\notag
\\
     & & \; \times \Big[ 1+ (\rho-\rho_c))^+ \int_{D}f(x)
         \cos^2\pi x_3 \, dx \Big]^{-1},
\label{mes}
\end{eqnarray}
where $ c^+ = \max\{ c,  0\}$ and $ \, c\wedge d = \min\{c,d\}$.
The first equation expresses the BRPF for the microscopic scale,
the second the BRPF for the macroscopic and third for the mesoscopic
scale.
Here, the parameters appeared in every scale are written in terms of
$\rho$ which appeared in microscopic version in Remark \ref{rho}.

{From} the mesoscopic scale (\ref{mes}), we can see the $x_3$-dependence
of the local density, which is a effect of type-III gBEC.
The term containing $(\rho - \rho_{\rm m})^+$ in the denominator
in (\ref{mac}) shows the $x_1, x_2$-dependence of the local density in
this scale, which is a effect of the usual BEC.
In this way, the qualitative difference between the two BEC phases is
obvious.
\end{remark}

\subsection{Microscopic and macroscopic distribution of constituent
bosons in algebraically anisotropic Casimir prism (BEAM)}\label{sub-sec-BEAM}
\noindent \textbf{(a) Casimir prism: one critical temperature and type-II gBEC.}
In this subsection, we consider the distribution of constituent bosons
in the Casimir prism (BEAM) \cite{PuZ} in order to compare it to the SLAB case \cite{BZ}.

To review the picture of BEC in the BEAM system \cite{BLP} or \cite{Zag2}, let us consider first the
expectation value of the particle density in the grand canonical ensemble for the prism
$\Lambda_L := L_1 \times L_2 \times L_3$, where $L_1=L^{\gamma}, L_2=L_3=L$. Then its volume is equal to
$V_L := |\Lambda_L|= L^{2+\gamma}$. Similar to the case of SLAB, for a fixed $\Delta >0$, we get the limiting
particle density
\[
      \bar\rho(\Delta) := \lim_{L\rightarrow\infty}
     \bar\rho(L, \Delta) = \frac{1}{(2 \pi)^3}\int_{\mathbb{R}^3}
  \frac{dk}{e^{\beta(\hbar^2 k^2/2m + \Delta)}-1}
  <  \bar\rho_c \ ,
\]
and $\lim_{\Delta\downarrow 0}\bar\rho(\Delta) = \bar\rho_c$, where
$\bar\rho_c$ is the same \textit{critical density} as in the SLAB case.

To make $\bar\rho(\Delta) > \bar\rho_c$, we must let $\Delta$ be
$L$-dependent: $\Delta =\Delta(L)$. Note that in this case, we have
\[
      \lim_{L\to\infty} \frac{1}{V_L}
          \sum_{\textbf{k}\in \mathbb{N}^3, k_2>1, k_3>1 }
         W(k, L, \Delta(L)) = \bar\rho_c
\]
if $\lim_{L\to\infty}\Delta(L) = 0$, see (\ref{A31}).
The remaining terms are
\[
       \frac{1}{V_L}
          \sum_{\textbf{k}\in \mathbb{N}^3, k_2 = k_3=1 }
         W(k, L, \Delta(L)) =
          \frac{1}{V_L}
          \Big\{ W({\bf 1}, L, \Delta(L)) +
       \sum_{\textbf{k}\in \mathbb{N}^3, k_1 >1, k_2 = k_3=1 }
         W(k, L, \Delta(L))\Big\}
\]
\begin{equation}
      =  \frac{1}{L^{2+\gamma}\beta\Delta(L)}
         + \frac{1}{L^2}O(L^{\gamma}\wedge \Delta(L)^{-1/2})\ ,
\label{avedsty2}
\end{equation}
see (\ref{A6}), where we put $L^{\gamma}$ instead of $L$.
Here the first term in (\ref{avedsty2}) is a contribution from the single-particle ground state,
and the second term is a contribution of countably many single-particle excited states.

Notice that for $\gamma < 2$ the $\lim_{L\rightarrow\infty} \bar\rho(\Delta(L)) =: \bar\rho  > \bar\rho_c$
implies $\Delta(L) = O(L^{-2-\gamma})= O(1/V_{L})$. Hence, only the first term in (\ref{avedsty2}) is non-zero in the limit,
that yields  the type-I (\textit{ground-state}) BEC.

To insure that $\lim_{L\rightarrow\infty} \bar\rho(\Delta(L)) =: \bar\rho  > \bar\rho_c$ for $\gamma = 2$,
one must put $\Delta(L) = O(L^{-4}) = O(1/V_{L})$. Then (\ref{avedsty2}) implies that not only ground-state but also
the contribution to the second term from each single-particle state is nonzero in thermodynamic limit.
Hence, for $\gamma = 2$ the gBEC in BEAM is of the type-II \cite{BLP}, \cite{Zag2}.

If $\gamma>2$, then to insure $\lim_{L\rightarrow\infty} \bar\rho(\Delta(L)) =: \bar\rho  > \bar\rho_c$
we have to put $\Delta(L) = O(L^{-4}) = O(1/V_{L}^{4/(2+\gamma)})$ \cite{BLP}. It is known that in this case
there is no macroscopic occupation of \textit{any} single-particle state, which corresponds to the type-III gBEC,
\cite{BLP}, \cite{Zag2}.

In the following we consider only the most interesting \textit{boundary case} $\gamma =2$, i.e. the case of the
type-II gBEC.

First, we examine the microscopic distribution of constituent particles in this kind of the Casimir prism (BEAM)
in the thermodynamic limit. We consider also $\nu_{\mu, \Lambda_L}$ and the BRPF (\ref{gf0}) for
$\Lambda =\Lambda_L$ and in the limit $L \rightarrow \infty$.

We introduce a function $\varphi$ on $ [0, \infty)$ defined by
\begin{equation}\label{phix}
      \varphi(x) := \sum_{n=1}^{\infty}\frac{1}{(2n-1)^2 -1 + x} \ .
\end{equation}
It is not difficult to show that $\varphi$ is decreasing and its
asymptotic behavior is
\[
  \varphi(x) = \begin{cases}
                     1/x + O(1) \quad & \mbox{for  small } \; x,  \\
                     \pi/(4\sqrt x)+O(1/x)\quad & \mbox{for large } \; x,
               \end{cases}
\]
(see Remark \ref{phiRuv}).
For the present system, we use the quantity
\[
    \tilde\kappa = \frac{16m}{\beta\pi^2\hbar^2}\lim_{L\to \infty}
        \varphi\big(2mL^4\Delta(L)/\pi^2\hbar^2\big) .
\]
We will see that each term of $\varphi$ reflects the contributions
of each state in
$ \{\, \phi_{(s_1,1,1)\Lambda} \, | \, s_1 : \rm{ odd} \, \} $,
in the proof of the theorem.
This indicates that the type II gBEC occurs in the system for
the case $\tilde\kappa \ne 0 $.

Now we can formulate the  following theorem:
\begin{thm}
If $\lim_{L\to \infty} \Delta(L) =\Delta_{\infty} >0 $,
then the BRPFs $\{ \, \nu_{\mu,\Lambda_L}\, \}_{L>0} $ converge weakly
to the BRPF $\nu_{\Delta_{\infty}, 0}$.
If $\lim_{L\to \infty} \Delta(L) = 0 $,
then the BRPFs $\{ \, \nu_{\mu,\Lambda_L}\, \}_{L>0} $ converge weakly
to the BRPF $\nu_{0, \tilde\kappa}$.
Here the BRPFs {\rm{:}} $\nu_{\Delta, 0}, \nu_{0, \kappa}$ (with $\kappa $ replaced by $ \tilde\kappa$),
coincide with those appeared in Theorem \ref{thmA}.
\label{thmD}
\end{thm}
\begin{remark}
As in  Remark \ref{rho}, we have
\[
        \rho = \tilde\kappa + K^{\Delta_{\infty}}(x,x)
\]
for the expected ``density" of the system.

We may also derive $L$-dependence of $\Delta(L)$ in order to give
a prescribed $\beta$ and $\rho$.
That is,

\noindent for $(\beta, \rho)$ in the region
$\rho \leqslant \rho_c$ (normal phase),
we may take $L$-independent $\Delta$
which is the unique solution of
\[
      \rho = \int_{\R^3}\frac{1}
     {e^{\beta(\hbar^2|p|^2/2m+\Delta)}-1} \,
        \frac{dp}{(2\pi)^3}
\]
and for $(\beta, \rho)$ in the region $\rho > \rho_c $
(type-II gBEC case),
$\Delta(L)$ is given by
\[
       \Delta(L) = \frac{\pi^2\hbar^2}{2mL^4}
 \varphi^{-1}\Big(\frac{(\rho- \rho_c)\beta\pi^2\hbar^2}{16}
\Big),
\]
where $\rho_c$ is the same one as in the SLAB case.
\end{remark}

\smallskip

Now let us consider the two type of macroscopic distribution of the
constituent particles of the Casimir prism by using two different scaling
arguments.

In the first scaling, $\Lambda_L$ looks like
$ R = \R \times [-1/2, 1/2]^2$
and the corresponding transformation is
\[
     T_R: Q(\Lambda_L) \ni \xi = \sum_j \delta_{x^{(j)}} \longmapsto
             \eta = \frac{1}{L^3} \sum_j
      \delta_{x^{(j)}/L } \in {\, \cal M}(R).
\]

Let us introduce the random field
\begin{equation}
        \mathscr{M}^{(R)}_{\mu, L} =
       \nu_{\mu, \Lambda_L} \circ (T_R)^{-1}.
\label{TR}
\end{equation}
We also define RF $\mathscr{M}^{(R)}_{a, b}$, which is characterised by the
generating functional
\[
          \int_{{\cal M}(R)}e^{-\langle f, \eta \rangle}
          \mathscr{M}^{(R)}_{a, b}(d\eta)
       =  \exp\Big[-a \int_{R} f(x) \, dx \Big]
\]
\[
       \times \Big[ 1+ b \int_{R}f(x)
         \big(\prod_{j=2}^3\cos^2\pi x_j \big) \, dx \Big]^{-1},
\]
for non-negative $f \in C_0(R)$.
We get the following result:
\begin{thm}
The random measure $ \mathscr{M}^{(R)}_{\mu, L} $ converges to
$ \mathscr{M}^{(R)}_{a, b} $
with
\[
      a = K^{\Delta_{\infty}}(x, x)
\]
and  $b =  \tilde \kappa$, weakly.
\label{thmE}
\end{thm}
\begin{remark}
 This distribution is a convex combination of ``deterministic states"
as in Remark \ref{remB1}.
\end{remark}
\begin{remark}
We have
\[
          \int_{{\cal M}(R)} \langle f, \eta \rangle
          \mathscr{M}^{(R)}_{a, b}(d\eta)
       =   \int_{R} \rho(x) f(x) \, dx,
\]
where
\[
       \rho (x) = K^{\Delta_{\infty}}(x,x) + \tilde\kappa
        \prod_{j=2}^3\cos^2\pi x_j
\]
is the ``density" of the system in the mesoscopic scale.
The second term of $\rho$ is proportional to the value at
$(0, x_2,x_3)$ of the scaled and squared low-lying eigenfunctions
$ \{\, \phi_{(s,0,0) \Lambda_L} \, \}_{ s\in \N }$.
\end{remark}

\bigskip

In the second scaling for Casimir prism, the system looks like
$ I=[-1/2,  1/2]$ in the limit.
The transformation for the scaling is
\[
     T_I: Q(\Lambda_L) \ni \xi = \sum_j \delta_{x^{(j)}} \longmapsto
             \eta = \frac{1}{L^4}\sum_j
    \delta_{x_1^{(j)}/L^2}  \in {\cal M}(I) .
\]
We introduce the RF
\begin{equation}
        \mathscr{M}^{(I)}_{\mu, L} = \nu_{\mu, \Lambda_L}
      \circ (T_I)^{-1}.
\label{MIT}
\end{equation}
We also define the RF $ \mathscr{M}^{(I)}$ on ${\cal M}(I)$,
which is characterized by the generating functional:
\begin{equation}\label{exD}
          \int_{{\cal M}(I)}e^{-\langle f, \eta \rangle}
          \mathscr{M}^{(I)}(d\eta)
       =  \frac{\exp\Big[- K^{\Delta_{\infty}}(x, x)
      \int_{I} f(u) \, du \Big]} {\Det [ 1+ fR]} \ ,
\end{equation}
where $R$ is the integral operator whose kernel is
\begin{equation}\label{Ruv}
R(u,v) := \frac{8m}{\beta\hbar^2\pi^2} \sum_{n=1}^{\infty} \frac{\cos n\pi(u-v) - \cos n\pi(1+u+v)}{n^2 + \alpha^2}
\end{equation}
\begin{equation*}
       = \frac{8m}{\beta\hbar^2\pi^2}
    \frac{\pi}{2\alpha} \frac{\big[\cosh \big(\pi\alpha(1-|u-v|)\big)
       -\cosh\big(\pi\alpha(u+v)\big)\big]}{\sinh\pi\alpha} \ ,
\end{equation*}
with
\[
        \alpha = \lim_{L\to\infty}\sqrt{\frac{2mL^4\Delta(L)}
        {\hbar^2\pi^2} -1},
\]
and $f \in C(I)$ is non-negative.
The second equality of (\ref{Ruv}) is derived by the Fourier series
expansion for $\cosh$. Note that $\alpha $ may be pure imaginary. We understand that $R=0$ for $\alpha = \infty$.
We get the following result:
\begin{thm}
The random measure $ \mathscr{M}^{(I)}_{\mu, L} $ converges to
$ \mathscr{M}^{(I)} $, weakly.
\label{thmF}
\end{thm}
\begin{remark}
We have
\[
          \int_{{\cal M}(I)} \langle f, \eta \rangle
          \mathscr{M}^{(I)}(d\eta)
       =   \int_{I} \rho(u) f(u) \, du ,
\]
where
\[
        \rho(u)  = K^{\Delta_{\infty}}(x,x) + R(u,u)
        = K^{\Delta_{\infty}}(x,x) + \tilde\kappa
        \frac{\cosh (\pi\alpha) -\cosh(2\pi\alpha u)}
           {\cosh(\pi\alpha)-1}
\]
is the expected ``density".
Note that
$\displaystyle \varphi(\alpha^2+1) =
\frac{\beta \hbar^2\pi^2}{16m}R(0,0), \,
R(0,0) =\tilde \kappa$
and
\[
           \varphi(x) = \frac{\pi}{4\sqrt{x-1}}
            \frac{\cosh \pi\sqrt{x-1}-1}{\sinh\pi\sqrt{x-1}}
\]
holds.
\label{phiRuv}
\end{remark}
\begin{remark}
In this case, the macroscopic ``density" has random distribution.
In fact, it follows from (\ref{exD}) that the density is a squared
Gaussian RF.
The operator $R$ consists of the contributions of low-lying
eigenfunctions $ \{\, \phi_{(s,0,0) \Lambda_L} \, \}_{ s\in \N }$.
This revived random, which is disappeared once in mesoscopic scale,
may be considered as the contribution of boundary effects through
the law-lying eigenfunctions.
This are strong fluctuations, which are typical in type-II gBEC, see e.g.  \cite{BLL}, \cite{PuZ}.
\end{remark}

\section{Proofs for the SLAB }
\subsection{Proof of Theorem\,\ref{thmA} }
The function $f$, which will appear in the generating functional
$ Z_{\Lambda}(f)$ is assumed to be non-negative continuous function on
$\R^3$ with compact support.
In the course of the thermodynamic limit, $f$ is regarded as a function
in $C_0(\Lambda_L)$ naturally.

Let $P_0$ be the orthogonal projection operator on $L^2(\Lambda_L)$ onto
$\mathscr{H}_0$, the closed subspace spanned by
$\{ \phi_{(s_1,s_2,1)\Lambda} \, | \, s_1,s_2 \in \N \}$, and $P_1$
the orthogonal projection onto $\mathscr{H}_1=\mathscr{H}_0^{\perp}$.
Let us recall that the function $K_{\Lambda}(x,y)$ is given by
(\ref{ODLROkernel}) with $L_1=L_2=Le^{\alpha L}, L_3=L$ and $L$-dependent
$\Delta = \Delta(L)$, though the dependence on
$\Delta( \, \cdot \, )$ is suppressed on the symbol.
In the following, we also suppress the $\Delta_{\infty}$-dependence
from $K^{\Delta_{\infty}}(x,x), K^{\Delta_{\infty}}_f$, etc., and we write $\Delta$ instead of
$ \Delta(L)$.
\begin{lem}
The following convergence of the operators holds in trace class:
\[
         \sqrt{1-e^{-f}}P_1K_{\Lambda}P_1\sqrt{1-e^{-f}}
    \longrightarrow  K_f
\]
\label{lem1}
\end{lem}
\noindent{\it Proof }: From (\ref{A3}),
\[
    P_1K_{\Lambda}P_1(x, y) = \sum_{s_1, s_2 =1}^{\infty}\sum_{s_3=2}^{\infty}
    \frac{\phi_{s,\, \Lambda}(x)\overline{\phi_{s, \,\Lambda}(y)}}
     {\exp\Big[\frac{\beta\hbar^2\pi^2}{2m}
     \Big( \frac{s_1^2-1+s_2^2-1}{L^2e^{2\alpha L}}+ \frac{s_3^2-1}{L^2}\Big)
      +\beta\Delta  \Big] - 1}
\]
\[
    \longrightarrow  \int_{\R^3}\frac{e^{2\pi ip\cdot(x-y)}}
     {\exp\Big[\frac{4\beta\hbar^2\pi^2}{2m}|p|^2 +
       \beta\Delta_{\infty}\Big]
        -1}\,dp = K(x, y)
\]
holds locally uniformly in $x, y$.
Since supp$\, f$ is compact, we have
\[
        \Tr [\sqrt{1-e^{-f}}P_1K_{\Lambda}P_1\sqrt{1-e^{-f}}]
        = \int_{\Lambda} P_1K_{\Lambda}P_1(x, x)(1-e^{-f(x)})\, dx
\]
\[
     \longrightarrow \int_{\R^3} K(x, x)(1-e^{-f(x)})\, dx
         = \Tr K_f.
\]
On the other hand,
\[
      \| \sqrt{1-e^{-f}}P_1K_{\Lambda}P_1\sqrt{1-e^{-f}}
    -  K_f \, \|
\]
\[
     = \sup_{\varphi, \psi \in L^2(\Lambda) \atop \|\varphi\|=\|\psi\|=1}
       (\sqrt{1-e^{-f}}\varphi, (P_1K_{\Lambda}P_1 -
       G(e^{\beta\Delta}-G)^{-1})\sqrt{1-e^{-f}} \psi)
\]
\[
      \leqslant \sup_{\varphi, \psi }\int_{\Lambda\times\Lambda}
       \sqrt{1-e^{-f(x)}}|\varphi(x)|\sqrt{1-e^{-f(y)}} |\psi(y)| \, dxdy
         \times   \sup_{\tilde x, \tilde y \in \supp f}
      |P_1K_{\Lambda}P_1(\tilde x, \tilde y) - K(\tilde x, \tilde y)|
\]
\[
      \leqslant \sup_{\varphi, \psi } \|1-e^{-f(x)}\|_1\|\varphi\|_2
       \|\psi\|_2 \sup_{\tilde x, \tilde y \in \supp f}
      |P_1K_{\Lambda}P_1(\tilde x, \tilde y) - K(\tilde x, \tilde y)|
      \longrightarrow 0
\]
holds.
Then, the Gr\"{u}mm convergence theorem \cite{Zag} yields the lemma.
\hfill $\square$

\begin{lem}
The following convergence of the operators holds in the trace-class topology:
\[
         P_0K_{\Lambda}P_0
    \longrightarrow  \kappa\sqrt{1-e^{-f}}(\sqrt{1-e^{-f}}, \cdot ) \ .
\]
Here the limit operator $\kappa\sqrt{1-e^{-f}}(\sqrt{1-e^{-f}}, \cdot )$ is a projector, which acts as
\[
   \psi \longmapsto (\kappa_1 + \kappa_2)( \sqrt{1-e^{-f}}, \psi) \sqrt{1-e^{-f}} \ , \ \kappa=\kappa_1 + \kappa_2 \ ,
\]
for $\psi \in L^2(\R^3)$.
\label{lem0}
\end{lem}
\noindent{\it Proof }: In the expression
\[
    P_0K_{\Lambda}P_0(x, y) = \sum_{s_1, s_2 =1}^{\infty}
    \frac{\phi_{(s_1,s_2,1),\, \Lambda}(x)
      \overline{\phi_{(s_1, s_2, 1), \,\Lambda}(y)}}
     {\exp\Big[\frac{\beta\hbar^2\pi^2}{2m}
      \frac{s_1^2-1+s_2^2-1}{L^2e^{2\alpha L}}
      +\beta\Delta  \Big] - 1},
\]
the numerator may be represented as
\begin{equation}
      \phi_{(s_1,s_2,1),\, \Lambda}(x)
    \overline{\phi_{(s_1, s_2, 1), \,\Lambda}(y)}
\label{phiphi}
\end{equation}
\[
   = \Big\{\prod_{j=1}^2 \frac{2}{Le^{\alpha L}}\Big[\frac{1-(-1)^{s_j}}{2} +
       (-1)^{s_j}\sin^2\frac{\pi s_j(x_j + y_j)}{2Le^{\alpha L}} -
     \sin^2\frac{\pi s_j(x_j - y_j)}{2Le^{\alpha L}} \Big]\Big\}
        \frac{2}{L} \cos\frac{\pi x_3}{L}\cos\frac{\pi y_3}{L}.
\]
Hence we get
\begin{equation}
    P_0K_{\Lambda}P_0(x, y) =  \sum_{s_1, s_2 \in \N, {\rm odd}}
       \frac{8}{L^3e^{2\alpha L}}
         \frac{\cos(\pi x_3/L)\cos(\pi y_3/L)}
     {\exp\Big[\frac{\beta\hbar^2\pi^2}{2m}
      \frac{s_1^2-1+s_2^2-1}{L^2e^{2\alpha L}}
      +\beta\Delta  \Big] - 1} + R,
\label{00}
\end{equation}
where
\[
   |R| \leqslant \sum_{s_1, s_2 \in \N}
      \frac{1}{L^3e^{2\alpha L}}
     \frac{C(s_1^2 + s_2^2)/L^2e^{2\alpha L}}
     {\exp\Big[\frac{\beta\hbar^2\pi^2}{2m}
      \frac{s_1^2-1+s_2^2-1}{L^2e^{2\alpha L}}
      +\beta\Delta  \Big] - 1},
\]
where $C$ depends only on $|x|$ and $|y|$.
It follows from (\ref{A2}) and (\ref{A5}) that
\[
    P_0K_{\Lambda}P_0(x, y) =  \Big(\frac{8}{\beta\Delta L^3e^{2\alpha L}}
                + \frac{m}{\beta\pi\hbar^2L}\log\big[(L^2e^{2\alpha L})
                 \wedge (\beta\Delta)^{-1}\big]\Big)
             \cos\frac{\pi x_3}{L}\cos\frac{\pi y_3}{L}
\]
\begin{equation}
        + O(L^{-1}) +O(L^{-5}e^{-4\alpha L}(\beta\Delta)^{-1})
        \longrightarrow \kappa_1 + \kappa_2
\label{kappa12}
\end{equation}
locally uniformly in $x, y$ when $L\to\infty$.
The first term of the 2nd member in the above equation corresponds to
$s=(1,1)$.
Now, the lemma can be proved by the argument similar to the proof of the previous lemma.

\hfill $\square$

By consequence the \textit{weak} convergence of the BRPFs is implied by the following convergence
of the generating functional:
\[
      \int_{Q(\R^3)}e^{-\langle f, \xi\rangle}\nu_{\mu, {\Lambda}}(d\xi)
           = \Det\big[1 + \sqrt{1-e^{-f}}
         (P_0K_{\Lambda}P_0 + P_1K_{\Lambda}P_1)\sqrt{1-e^{-f}}\,\big]^{-1}
\]
\[
    \longrightarrow \Det\big[1+\kappa \sqrt{1-e^{-f}}(\sqrt{1-e^{-f}},
             + K_f \,\big]^{-1}
\]
\[
    = \Det[1+ K_f ]^{-1}
          \Det\big[1+ \kappa \sqrt{1-e^{-f}} (\sqrt{1-e^{-f}},
     (1+ K_f )^{-1} \big]^{-1}
\]
\[
    = \Det[1+ K_f]^{-1}
       \big\{1+ \kappa (\sqrt{1-e^{-f}},
     (1+ K_f )^{-1}\sqrt{1-e^{-f}} \, )\big\}^{-1}
\]
\[
     =   \int_{Q(\R^3)}e^{-\langle f, \xi\rangle}\nu_{\kappa}(d\xi) \ .
\]
Here we used Lemma \ref{lem1}, Lemma \ref{lem0} and
\begin{equation*}
\Det[1+A+B] = \Det[1+B] \ \Det[1+A(1+B)^{-1}] \ .
\end{equation*}
One has also to notice that projector: $ \, \kappa \sqrt{1-e^{-f}} (\sqrt{1-e^{-f}}, \cdot )\, $, is one-dimensional
operator.
\subsection{Proof of Theorem\,\ref{thmB} }
Let $P$ be the orthogonal projection operator on $L^2(\Lambda_L)$ onto
its one-dimensional subspace $ \C \phi_{{\bf 1}\Lambda_L}$, and $Q$
the orthogonal projection onto $ \big(\C \phi_{{\bf 1}\Lambda_L}\big)^{\perp}$.

Put
\[
    f_L(x) = L^{-3}e^{-2\alpha L}f
        \big( x_1/Le^{\alpha L}, x_2/Le^{\alpha L}\big)
\]
and
\[
    f^{(L)}(x) = L^3e^{2\alpha L}\big(1-e^{-f(x)/L^3e^{2\alpha L}}\big)
\]
for bounded measurable non-negative function $f$ on
$S=[-1/2, 1/2]^2$.
The functions $f_L, f^{(L)}$ are defined on  $\Lambda_L$ and on
$S$, respectively.
Note that $f^{(L)}(x) \uparrow f(x) $.
Then we have
\begin{equation}
              \langle f_L, \xi \rangle = \langle f, \eta \rangle
           =   \langle f, T_S\xi \rangle.
\label{xieta}
\end{equation}

{From}  (\ref{gf0}), (\ref{gfD}), (\ref{xieta}) and  (\ref{T}), we have
\[
        \int_{{\cal M}(S)}
         e^{-\langle f, \eta\rangle} \mathscr{M}_{\mu, L}^{(S)}
        (d\eta)
         = \int_{Q(\Lambda)}
         e^{-\langle f_L, \xi\rangle} \nu_{\mu, \Lambda_L}(d\xi)
      = \Det[1+K_{\Lambda_L}(1-e^{-f_L})]^{-1}.
\]
By virtue of the \textit{Feshbach formula}  we get
\[
        \Det[1+K_{\Lambda}(1-e^{-f_L})]^{-1}
      = \Det[1 + (P+Q)K_{\Lambda}(1-e^{-f_L})(P+Q)\,]^{-1}
\]
\[
      = \Det\big[1 + PK_{\Lambda}(1-e^{-f_L})P - PK_{\Lambda}(1-e^{-f_L})Q
    \big( 1+ QK_{\Lambda}(1-e^{-f_L})Q\big)^{-1} QK_{\Lambda}(1-e^{-f_L})P
        \, \big]^{-1}
\]
\begin{equation}
       \times \Det[1 + QK_{\Lambda}(1-e^{-f_L})Q]^{-1}.
\label{SDD}
\end{equation}

Note that
\[
        PK_{\Lambda}P =(e^{\beta\Delta}-1)^{-1}P,  \quad
        \| PK_{\Lambda}P\| \leqslant (\beta\Delta)^{-1},
\]
\[
         \|QK_{\Lambda}Q\| =
   \Big(\exp\Big[\frac{\beta \hbar^2\pi^2(4-1)}{2mL^2e^{2\alpha L}}
    + \beta\Delta\Big]-1 \Big)^{-1} = O\big(L^2e^{2\alpha L}\wedge
       (\beta\Delta)^{-1}\big)
\]
and
\[
     \| 1-e^{-f_L} \| \leqslant \|f\|_{\infty}/L^3e^{2\alpha L}, \quad
     \|QK_{\Lambda}(1-e^{-f_L})Q\| =O(L^{-1}).
\]
It follows that
\[
     \| PK_{\Lambda}(1-e^{-f_L})Q
    \big( 1+ QK_{\Lambda}(1-e^{-f_L})Q\big)^{-1} QK_{\Lambda}(1-e^{-f_L})P\|
\]
\[
            \leqslant \frac{1}{\beta\Delta}
       \frac{\|f\|_{\infty}}{L^3e^{2\alpha L}}
        O\big(L^{-1}\big)
                =O(L^{-1}).
\]
Here, we recall the assumption of boundedness of
$(\Delta L^3e^{2\alpha L})^{-1}$ and that
$QK_{\Lambda}(1-e^{-f_L})Q$ is positive.
The convergence of the one-dimensional operator
\[
       PK_{\Lambda}(1-e^{-f_L})P = \frac{(\phi_{{\bf 1}, \Lambda},
       (1-e^{-f_L})\phi_{{\bf 1}, \Lambda})}{e^{\beta\Delta} - 1}P
\]
\[
     =   \frac{4}{( e^{\beta\Delta} - 1) L^3e^{2\alpha L}}
         \int_{S}f^{(L)}(x)
   \big(\prod_{j=1}^2\cos^2\pi x_j\big) \, dx
     \, P
\]
\[
     \longrightarrow \Big(\lim_{L\to \infty}
      \frac{4}{\beta\Delta L^3e^{2\alpha L}}\Big)
      \int_{S}f(x)
      \big(\prod_{j=1}^2\cos^2\pi x_j\big) \, dx \, P
\]
holds.
Note that the trace norm is identical to the operator norm for
a one-dimensional operator.
Thus we get
\[
      \Det\big[1 + PK_{\Lambda}(1-e^{-f_L})P - PK_{\Lambda}(1-e^{-f_L})Q
    \big( 1+ QK_{\Lambda}(1-e^{-f_L})Q\big)^{-1} QK_{\Lambda}(1-e^{-f_L})P
        \, \big]
\]
\begin{equation}
       \longrightarrow 1 + \Big(\lim_{L\to \infty}
       \frac{4}{\beta\Delta L^3e^{2\alpha L}}\Big)
      \int_{S}f(x)
     \big(\prod_{j=1}^2 \cos^2\pi x_j \big) \, dx.
\label{Sb}
\end{equation}

We remark here that the function $ \prod_{j=1}^2 \cos^2\pi x_j $ is a reminiscent of the
\textit{ground-state} wave function.

Next, let us consider the operator $QK_{\Lambda}(1-e^{-f})Q$.
To simplify the notations, we introduce the following symbols:
\begin{equation}\label{symb1}
a_{s_1, s_2}^{(L)} = \int_{S}f^{(L)}(u)\big[\prod_{j=1}^2\cos(2\pi s_ju_j)\big]\, du \ ,
\end{equation}
the corresponding limit:
\begin{equation}\label{symb2}
a_{s_1, s_2} = \int_{S}f(u)\big[\prod_{j=1}^2\cos(2\pi s_ju_j)\big]\, du \ ,
\end{equation}
for $s=(s_1, s_2) \in \bar\N^2$, where $\bar \N = \{0\} \cup \N$,
and the symbol:
\begin{equation}\label{symb3}
W(s, L) = \Big(\exp\Big[\frac{\beta \hbar^2\pi^2}{2m} \Big( \frac{s_1^2-1+s_2^2-1}{L^2e^{2\alpha L}} +
\frac{s_3^2-1}{L^2} \Big) + \beta\Delta\Big]-1 \Big)^{-1}
\end{equation}
for $ s = (s_1, s_2, s_3) \in \N^3$.

Then argument based on the \textit{Riemann-Lebesgue} lemma yields the following result:
\begin{lem}\label{lem11}
For every $s = (s_1, s_2)  \in\bar\N^2 $,
\[
         |a_s|, \; |a_s^{(L)}| \leqslant \|f\|_{\infty}, \quad
         |a_s -a_s^{(L)}| \leqslant \|f\|_{\infty}^2/2L^3e^{2\alpha L}
\]
holds. And $a_s^{(L)}$ converges to 0, i.e.,
\[
      \forall \epsilon >0, \exists N_0 \in \N, \exists L_0>0:
      s_1+ s_2 > N_0, \, L >L_0 \Longrightarrow
           |a_s^{(L)}| < \epsilon.
\]
\end{lem}

Returning back to the proof of Theorem \ref{thmB} note that in the expression
\[
     \Tr [QK_{\Lambda}(1-e^{-f_L})Q] = \sum_{s\in\N^3 \atop s \ne {\bf 1}}
     W(s, L) \int_{\Lambda}|\phi_{s, \Lambda}(x)|^2
         (1-e^{-f_L(x)})\,dx
\]
\[
         = \sum_{s\in\N^3 \atop s \ne {\bf 1}}\frac{W(s, L)}{L^3e^{2\alpha L}}
      \int_{S} f^{(L)}(u) \prod_{j=1}^2
        \big(1-(-1)^{s_j}\cos 2\pi s_ju_j\big)\,du
\]
\[
     =: \, a_{(0,0)}^{(L)}\sum_{s\in\N^3 \atop s \ne {\bf 1}}
       \frac{W(s, L)}{L^3e^{2\alpha L}} + R_L \ ,
\]
eqs.(\ref{A1}), (\ref{A4}) yield
\[
       \sum_{s\in\N^3 \atop s \ne {\bf 1}}
       \frac{W(s, L)}{L^3e^{2\alpha L}}
      = \sum_{s\in\N^3 \atop s \ne {\bf 1}, s_3 =1} \ldots +
         \sum_{s\in\N^3 \atop s_3\ne 1}\ldots
\]
\[    \longrightarrow
       \lim_{L\to \infty}\Big[\frac{\pi}{4\beta \hbar^2\pi^2/2m}
       \frac{1}{L}\log\big(L^2e^{2\alpha L}\wedge (\beta\Delta)^{-1}\big)\Big]
\]
\[
       + \frac{1}{8}\int_{\R^3}\frac{dp}{ \exp\big[\beta \hbar^2\pi^2 |p|^2/2m
           + \beta\Delta_{(L=\infty)}\big]-1}
\]
as $L\to \infty $, and we shall show that $\lim_{L\to \infty}R_L =0$ in Lemma \ref{lem12} below.

Then together with observation: $\| QK_{\Lambda}(1-e^{-f_L})Q \| \to 0$ as $L \to \infty$, we get
\[
     \Det\big[ 1+ QK_{\Lambda}(1-e^{-f_L})Q \big]
       = e^{\Tr[QK_{\Lambda}(1-e^{-f_L})Q] + o(1)}
\]
\[
        \longrightarrow \exp\Big[ a_{(0,0)}\Big( K(x,x) +
            \frac{m}{\beta 2\pi \hbar^2}\lim_{L\to \infty}
       \frac{1}{L}\log\big(L^2e^{2\alpha L}\wedge (\beta\Delta)^{-1}\big)
       \Big)\Big].
\]
\begin{equation}
      =e^{a\int_Sf(u)\, du}
\label{Sa}
\end{equation}
As a consequence we obtain from (\ref{SDD}), (\ref{Sb}) and (\ref{Sa}) that
\[
      \int_{{\cal M}(S)}e^{-\langle f, \eta \rangle}
         \mathscr{M}^{(S)}_{\mu, L}(d\eta)
     \to e^{a\int_Sf(u)\, du}
      \Big(1+b\int_Sf(u)\big(\prod_{j=1}^2 \cos\pi u_j\big)\,
             du\Big)^{-1}
\]
\[
     =  \int_{{\cal M}(S)}e^{-\langle f, \eta \rangle}
      \mathscr{M}^{(S)}_{a, b}(d\eta) \ .
\]
This proves the Theorem \ref{thmB}.  \hfill $\square$

Now we return to the promised Lemma \ref{lem12}:
\begin{lem}\label{lem12}
\[
         \lim_{L\to \infty}R_L =0
\]
\end{lem}
{\sl Proof : } With help of (\ref{symb1}) and (\ref{symb3}) we can re-write $R_L$ as the sum of three terms
$R_L = R_L^{(1)} + R_L^{(2)} + R_L^{(3)} $, where
\[
      R_L^{(1)} = \sum_{s\in\N^3, s\ne {\bf 1}}(-1)^{s_1 + 1}
      a_{(s_1,0)}^{L} \frac{W(s,L)}{L^3e^{2\alpha L}}, \quad
      R_L^{(2)} = \sum_{s\in\N^3, s\ne {\bf 1}}(-1)^{s_2 + 1}
      a_{(0,s_2)}^{L} \frac{W(s,L)}{L^3e^{2\alpha L}}
\]
and
\[
       R_L^{(3)} = \sum_{s\in\N^3, s\ne {\bf 1}}(-1)^{s_1 + s_2}
      a_{(s_1,s_2)}^{L} \frac{W(s,L)}{L^3e^{2\alpha L}}.
\]
For an arbitrary but fixed $\epsilon >0$, let $L_0, N_0$ be the numbers which
appeared in Lemma \ref{lem11}.
Suppose $L>L_0$. Then, we have
\[
       |R_L^{(1)}| \leqslant \sum_{s\in\N^3, s\ne {\bf 1}}
      |a_{(s_1,0)}^{L}| \frac{W(s,L)}{L^3e^{2\alpha L}}
\]
\[
       \leqslant |a_{(1, 0)}|\sum_{s_2,s_3 \in\N, s_2+s_3 \geqslant 3}
      \frac{W((1, s_2,s_3),L)}{L^3e^{2\alpha L}}
      + \sum_{s_1=2}^{N_0} |a_{(s_1,0)}^{L}| \sum_{s_2,s_3 \in\N}
        \frac{W(s,L)}{L^3e^{2\alpha L}}
\]
\[
        + \sum_{s_1=N_0+1}^{\infty} |a_{(s_1,0)}^{L}| \sum_{s_2,s_3 \in\N}
        \frac{W(s,L)}{L^3e^{2\alpha L}}
\]
\[        \leqslant \|f\|_{\infty} \frac{1}{Le^{\alpha L}}
            \big[ O(e^{\alpha L}) + O(\log L) \big]
          + (N_0-1)\|f\|_{\infty} \Big\{ \frac{1}{Le^{\alpha L}}
          \big[ O(e^{\alpha L}) + O(\log L) \big]
               + \frac{W((2,1,1),L)}{L^3e^{2\alpha L}} \Big\}
\]
\[
               + \epsilon \sum_{s\in\N^3, s\ne {\bf 1}}
            \frac{W(s,L)}{L^3e^{2\alpha L}}
\]
\[
           = O(N_0\log L/L) + O(\epsilon).
\]
We have used (\ref{A7}) at the third inequality and
(\ref{A4}) at the last line.

Hence by taking $L$ large enough, we can make $R_L^{(1)}$ arbitrarily small.
Thus we get $  \lim_{L\to \infty}R_L^{(1)} =0$.

We get $  \lim_{L\to \infty}R_L^{(2)} =0$ in the same way.
For $  \lim_{L\to \infty}R_L^{(3)} =0 $, we proceed in a similar manner, i.e.,
\[
       |R_L^{(3)}| \leqslant \sum_{s\in\N^3, s\ne {\bf 1}}
      |a_{(s_1,s_2)}^{L}|\frac{W(s,L)}{L^3e^{2\alpha L}}
\]
\[
       \leqslant |a_{(1, 1)}|\sum_{s_3=2}^{\infty}
        \frac{W((1,1,s_3),L)}{L^3e^{2\alpha L}}
       + \sum_{3\leqslant s_1 +s_2 \leqslant N_0} \sum_{s_3=1}^{\infty}
         |a_{(s_1,s_2)}^{L}| \frac{W((s_1, s_2, s_3), L)}{L^3e^{2\alpha L}}
\]
\[
        + \sum_{s_1+s_2 >N_0 }^{\infty}
        |a_{(s_1, s_2)}^{L}| \sum_{s_3 =1}^{\infty}
        \frac{W(s,L)}{L^3e^{2\alpha L}}
\]
\[
       \leqslant \|f\|_{\infty} \sum_{s_3=2}^{\infty}
        \frac{W((1,1,s_3),L)}{L^3e^{2\alpha L}}
       + \Big(\frac{N_0(N_0-1)}{2}-1\Big) \|f\|_{\infty}
           \sum_{s_3=1}^{\infty} \frac{W((2,1,s_3), L)}{L^3e^{2\alpha L}}
\]
\[
        + \epsilon \sum_{s\in\N^3, s \ne {\bf 1} }
         \frac{W(s,L)}{L^3e^{2\alpha L}}
\]
\[
           = O(N_0^2/L) + O(\epsilon),
\]
where we have used (\ref{symb2}) and (\ref{A6}) in the last line. \hfill $\square$

\subsection{Proof of Theorem\,\ref{thmC} }
Let
$D_L=[-e^{\alpha L}/2,  e^{\alpha L}/2]^2 \times [-1/2, 1/2]$
and
$D = {\Bbb R}^2\times [-1/2, 1/2]$.  Put
\[
    f_L(x) = L^{-3}f\big( x/L\big)
\]
and
\[
    f^{(L)}(x) = L^3\big(1-e^{-f(x)/L^3}\big)
\]
for non-negative $f \in C_0(D)$. For large $L>0$ we can naturally consider $f \in C_0(D_L)$.
The functions $f_L, f^{(L)}$ are defined on  $\Lambda_L$ and on $D$, respectively.
Note that $f^{(L)}(x) \uparrow f(x)$. Then by definition (\ref{T_D}) we have
\begin{equation}\label{Dxieta}
\langle f_L, \xi \rangle =  \langle f, T_D \, \xi \rangle \ .
\end{equation}

Recall that  $P_0$ be the orthogonal projection operator on
$L^2(\Lambda_L)$ onto $\mathscr{H}_0$ the closed subspace
spanned by
$\{ \phi_{(s_1,s_2,1)\Lambda} \, | \, s_1, s_2 \in \N \}$,
and $Q_1$ the orthogonal projection onto
$\mathscr{H}_1=\mathscr{H}_0^{\perp}$.

{From} (\ref{gf0}), (\ref{gfD}), (\ref{Dxieta}) and (\ref{TD}), we deduce that
\[
        \int_{{\cal M}(D)}
      e^{-\langle f, \eta\rangle} {\mathscr{M}}_{\mu, L}^{(D)}
        (d\eta)
         = \int_{Q(\Lambda)}
       e^{-\langle f_L, \xi\rangle} \nu_{\mu, \Lambda_L}(d\xi)
\]
\[
      = \Det[1+ \sqrt{1-e^{-f_L}}K_{\Lambda_L}\sqrt{1-e^{-f_L}}]^{-1}
\]
\[      = \Det[1 + \sqrt{1-e^{-f_L}}(P_0K_{\Lambda_L}P_0 + P_1K_{\Lambda_L}P_1)
          \sqrt{1-e^{-f_L}} \,]^{-1}
\]
\[
      = \Det\big[1 + \sqrt{1-e^{-f_L}}P_1K_{\Lambda_L}P_1 \sqrt{1-e^{-f_L}}\big]^{-1}
\]
\[
       \times \Det[1 + \sqrt{1-e^{-f_L}}P_0K_{\Lambda_L}P_0\sqrt{1-e^{-f_L}}
              (1 + \sqrt{1-e^{-f_L}}P_1K_{\Lambda_L}P_1 \sqrt{1-e^{-f_L}})^{-1}]^{-1} \ .
\]

We skip the proof of the following obvious operator-norm estimate:
\begin{lem}\label{lem31}
\[
    \| \sqrt{1-e^{-f_L}}P_1K_{\Lambda_L}P_1\sqrt{1-e^{-f_L}}\|
       =O(L^{-1}).
\]
\end{lem}

On the other hand we need the proof of the following limit:
\begin{lem}\label{lem32}
\begin{equation}\label{lim32}
    \Tr [\sqrt{1-e^{-f_L}}P_1K_{\Lambda_L}P_1\sqrt{1-e^{-f_L}}]
    \xrightarrow[L\to \infty]{} K(x,x)\int_Df(x)\, dx
\end{equation}
\end{lem}
{\sl Proof : } By virtue of
\[
  2 \sin^2\Big(\frac{\pi s_j}{2} + \frac{\pi s_j}{L_j}Ly_j\Big)
         = 1-(-1)^{s_j}\cos \Big(\frac{2\pi s_j}{L_j}Ly_j\Big),
\]
we get the following representation for the left hand side of (\ref{lim32}):
\[
       \sum_{s \in \N^3, s_3 \ne 1}
       \frac{8}{L^{3}e^{2\alpha L}}
         \frac{1}
         {\exp\Big[\frac{\beta\hbar^2\pi^2}{2m}
      \sum_{j=1}^3\frac{s_j^2-1}{L_j^2}+\beta\Delta  \Big] - 1}
      \int_Df^{(L)}(y)\prod_{j=1}^3
        \sin^2\Big(\frac{\pi s_j}{2} + \frac{\pi s_j}{L_j}Ly_j\Big)\,dy
\]
\begin{equation}\label{321}
        = \sum_{s \in \N^3, s_3 \ne 1}
       \frac{1}{L^{3}e^{2\alpha L}}
         \frac{\int_Df^{(L)}(y)\, dy}{\exp\Big[\frac{\beta\hbar^2\pi^2}{2m}
      \sum_{j=1}^3\frac{s_j^2-1}{L_j^2}+\beta\Delta  \Big] - 1} + r_L \ ,
\end{equation}
where  $L_1 =L_2=Le^{\alpha L}, L_3=L$ and
\begin{equation}\label{r_L}
r_L := \sum_{s \in \N^3, s_3 \ne 1}
\frac{1}{L^{3}e^{2\alpha L}} \frac{1}{\exp\Big[\frac{\beta\hbar^2\pi^2}{2m}
\sum_{j=1}^3\frac{s_j^2-1}{L_j^2}+\beta\Delta  \Big] - 1} \ \times
\end{equation}
\[
      \times \int_Df^{(L)}(y)\Big[
             - \sum_{j=1}^3(-1)^{s_j}\cos\Big(\frac{2\pi s_j}{L_j}Ly_j\Big)
             + \sum_{1\leqslant j<k \leqslant 3}(-1)^{s_j+s_k}
                \cos\Big(\frac{2\pi s_j}{L_j}Ly_j\Big)
               \cos\Big(\frac{2\pi s_k}{L_j}Ly_k\Big)
\]
\[             -  \prod_{j=1}^3(-1)^{s_j}\cos\Big(\frac{2\pi s_j}{L_j}Ly_j
              \Big) \Big]\, dy.
\]
Now, let us put
\[
     \Psi_L(p)
    = \begin{cases}
          \frac{1}{\exp\Big[\frac{\beta\hbar^2\pi^2}{2m}
     \sum_{j=1}^3\frac{s_j^2-1}{L_j^2}+\beta\Delta  \Big] - 1}
            & \quad \mbox{ for } p_j
            \in \Big( \frac{s_j-1}{L_j}, \frac{s_j}{L_j}\Big],
            \; j= 1,2,3, \; s\in \N^3,\, s_3 \ne 1         \\
          0 & \quad \mbox{ for } p_3 \in \Big( 0, \frac{1}{L_3}\Big]
       \end{cases}
\]
and
\[
         a_L(p) = \int_Df^{(L)}(y)\Big[
             - \sum_{j=1}^3(-1)^{s_j}\cos\Big(\frac{2\pi s_j}{L_j}Ly_j\Big)
             + \sum_{1\leqslant j<k \leqslant 3}(-1)^{s_j+s_k}
                \cos\Big(\frac{2\pi s_j}{L_j}Ly_j\Big)
               \cos\Big(\frac{2\pi s_k}{L_j}Ly_k\Big)
\]
\[
             -  \prod_{j=1}^3(-1)^{s_j}\cos\Big(\frac{2\pi s_j}{L_j}Ly_j\Big)
               \Big]\, dy
               \quad \mbox{ for } p_j
               =\Big( \frac{s_j-1}{L_j}, \frac{s_j}{L_j}\Big],
               \; j= 1,2,3, \; s\in \N^3.
\]
Then we have
\[
          |\Psi_L(p)| \leqslant \Phi(p) \equiv
          \frac{1}{e^{\beta\hbar^2\pi^2|p|^2/4m}-1},
          \quad \Phi \in L^1\big((0, \infty)^3\big),
\]
because
\[
       \sum_{j=1}^3\frac{s_j^2-1}{L_j^2} \geqslant
    \sum_{j=1}^3\frac{s_j^2}{2 L_j^2} \qquad \mbox{for } \quad
      s \in \N^3, s_3 \ne 1.
\]
We also have $ \lim_{L \to \infty}a_L(p) = 0 \; a.e.$ by the Riemann-Lebesgue lemma, and
\[
      \lim_{L\to \infty}\Psi_L(p) = \frac{1}{e^{\beta\hbar^2\pi^2|p|^2/2m
       + \beta\Delta}-1} \quad a.e..
\]
Therefore, the dominated convergence theorem yields for (\ref{r_L})
\[
      \lim_{L\to \infty} r_L = \lim_{L\to \infty}\int_{(0, \infty)^3}
        a_L(p)\Psi_L(p) \, dp = 0 \ ,
\]
and for the first term of the left hand side of eq.(\ref{321})
\[
     \int_Df^{(L)}(y)\, dy \int_{[0, \infty)^3}\Psi_L(p)\, dp
      \longrightarrow \int_Df(y)\, dy\int_{[0, \infty)^3}
             \frac{dp}{e^{\beta\hbar^2\pi^2|p|^2/2m
       + \beta\Delta}-1} \ . \qquad
\]
\hfill $\square$
\begin{remark}\label{DetP1}
From the above two lemmata we obtain that
\[
        \lim_{L\to\infty} \Det\big[1 + \sqrt{1-e^{-f_L}}P_1K_{\Lambda_L}P_1
         \sqrt{1-e^{-f_L}}\big]
       = e^{ K(x, x){\textstyle \int _D}f(y)\,dy} \ .
\]
We refer to (\ref{Kxy}) for definition of $K(x, y)$.
\end{remark}
\begin{lem}\label{lem33}
\[
   \lim_{L\to \infty} \Det[1 + \sqrt{1-e^{-f_L}}P_0K_{\Lambda_L}P_0
    \sqrt{1-e^{-f_L}}
       (1 + \sqrt{1-e^{-f_L}}P_1K_{\Lambda_L}P_1 \sqrt{1-e^{-f_L}})^{-1}]
\]
\[
       = 1+ \kappa \int _Df(x)\cos^2\pi x_3\,dx
\]
\end{lem}
{\sl Proof : } Let $U_L: L^2(\Lambda_L) \to L^2(D_L)$ be the unitary operator given by
$(U_Lg)(x) = L^{3/2}g(Lx)$ for $x\in D_L$.
Then, the transformation law of integral kernels is given by
\[
        (U_LKU_L^{-1})(x, y)=L^3K(Lx,Ly)
\]
for $x, y \in D_L$.
Together with eq.(\ref{phiphi}), it follows that
\[
    L^{-3}(U_LP_0K_{\Lambda_L}P_0U_L^{-1})(x, y)
     = P_0K_{\Lambda}P_0(Lx, Ly)
\]
\begin{equation}\label{00'}
       =:  \sum_{s_1, s_2 \in \N, {\rm odd}}
       \frac{8L^{-3}e^{-2\alpha L}\cos(\pi x_3)\cos(\pi y_3)}
     {\exp\Big[\frac{\beta\hbar^2\pi^2}{2m}
      \frac{s_1^2-1+s_2^2-1}{L^2e^{2\alpha L}}
      +\beta\Delta  \Big] - 1} + \varrho_L \ ,
\end{equation}
with
\[
   |\varrho_L| \leqslant \sum_{s_1, s_2 \in \N}
      \frac{C}{L^3e^{2\alpha L}}
     \frac{\big(L^2(|x|^2 +|y|^2)(s_1^2 + s_2^2)/L^2e^{2\alpha L}\big)^{1/4}}
     {\exp\Big[\frac{\beta\hbar^2\pi^2}{2m}
      \frac{s_1^2-1+s_2^2-1}{L^2e^{2\alpha L}}
      +\beta\Delta  \Big] - 1}
\]
\[
     = (|x|^2 +|y|^2)O(L^{-1/2})
\]
where $C$ is a numerical constant.
We have used $\sin^2x \leqslant C'|x|^{1/2}$ and (\ref{A501}).
By (\ref{A2}), we get the convergence
\[
     L^{-3}(U_LP_0K_{\Lambda_L}P_0U_L^{-1})(x, y) \longrightarrow
      \kappa\cos\pi x_3\cos\pi y_3
\]
uniformly in $x, y$ on compacta.
Then, by the same argument as in Lemma \ref{lem1}, we get
\[
       U_L\sqrt{1-e^{-f_L}}P_0K_{\Lambda}P_0\sqrt{1-e^{-f_L}}U_L^{-1}
     = \sqrt{f^{(L)}}L^{-3}U_LP_0K_{\Lambda_L}P_0U_L^{-1}
        \sqrt{f^{(L)}}
\]
\[
          \longrightarrow \kappa \sqrt f\cos\pi x_3( \sqrt f\cos\pi x_3,
\]
in trace class. Then the lemma follows from Lemma \ref{lem31} and the fact
that a unitary transformation conserves the Fredholm determinant.  \hfill$ \square$

Remark \ref{DetP1} and Lemma \ref{lem33}, with arguments developed at the beginning of this subsection,
imply the proof of Theorem\,\ref{thmC}.

\section{Proofs for the Casimir prism (BEAM)}
\subsection{Proof of Theorem\,\ref{thmD}}
Recall that below the function $f \in C_0(\R^3)$ is non-negative and we consider
$f \in C_0(\Lambda_L)$ for restriction $\Lambda_L\subset \R^3$.

Let $Q_0$ be the orthogonal projection operator on $L^2(\Lambda_L)$ onto
its close subspace $\mathscr{K}_0$ spanned by
$ \{ \phi_{(s_1,1,1)\Lambda_L} \, | \, s_1 \in \N \, \}$,
and $Q_1$ the orthogonal projection onto
$ \mathscr{K}_1 = \big(\mathscr{K}_0\big)^{\perp}$.
\begin{lem}\label{lem41}
The following limit:
\[
         \sqrt{1-e^{-f}}Q_1K_{\Lambda}Q_1\sqrt{1-e^{-f}} \longrightarrow  K_f.
\]
is true in the trace-class topology $\|\cdot\|_1$.
\end{lem}
The proof of the lemma follows the lines reasoning of Lemma \ref{lem1}, using (\ref{A31}) instead of (\ref{A3}).
\begin{lem}\label{lem42}
One has the limit:
\begin{equation*}
   \|\cdot\|_1 - \lim Q_0K_{\Lambda}Q_0 = \tilde\kappa\sqrt{1-e^{-f}}(\sqrt{1-e^{-f}}, \ ,
\end{equation*}
where $\tilde\kappa$ is defined by (\ref{kappa-tilda}).
\end{lem}
{\sl Proof : } Similar to eq.(\ref{00}) in the proof of Lemma \ref{lem0},
we get

\begin{equation}
    Q_0K_{\Lambda}Q_0(x, y) =:  \sum_{s_1 \in \N, {\rm odd}}
       \frac{8}{L^4}\frac{\cos(\pi x_2/L)\cos(\pi y_2/L)
            \cos(\pi x_3/L)\cos(\pi y_3/L)}
     {\exp\Big[\frac{\beta\hbar^2\pi^2}{2m}
      \frac{s_1^2-1}{L^4}
      +\beta\Delta  \Big] - 1} + \widehat{R}_L \ ,
\end{equation}
where
\[
   |\widehat{R}_L| \leqslant \sum_{s_1 \in \N}
      \frac{C}{L^4}
     \frac{s_1^2 /L^4}
     {\exp\Big[\frac{\beta\hbar^2\pi^2}{2m}
      \frac{s_1^2-1}{L^4}
      +\beta\Delta  \Big] - 1} = O(L^{-7}(\beta\Delta)^{-1}) +O(L^{-2}) \ ,
\]
where $C$ depends only on $|x|$ and $|y|$.
We used (\ref{A51}) in the last estimates.
Then, by virtue of (\ref{A21}), we get that
\begin{equation}\label{kappa-tilda}
       Q_0K_{\Lambda}Q_0(x, y) \longrightarrow
       \frac{16m}{\beta\pi^2\hbar^2}\lim_{L\to \infty}
        \varphi\big(2mL^4\Delta/\pi^2\hbar^2\big) = \tilde\kappa
\end{equation}
converges uniformly in $x, y$ on supp$\,f$.
The rest of the proof is similar to that of Lemma \ref{lem1}.  \hfill $\square$

Now, taking into account Lemma \ref{lem41} and Lemma \ref{lem42}, the Theorem \ref{thmD} can
be proved along the same lines of reasoning as Theorem \ref{thmA}.
\subsection{Proof of Theorem\,\ref{thmE}}
Let $ R := \R \times [-1/2, 1/2]^2$ and $R_L := [-L/2, L/2] \times [-1/2, 1/2]^2$. Put
\[
    f_L(x) = L^{-3}f(x/L) \ ,
\]
and
\[
    f^{(L)}(x) = L^3\big(1-e^{-f(x)/L^3}\big) \ ,
\]
for bounded measurable non-negative function $f$ on ${R}$.
Then, the functions $f_L, f^{(L)}$ are defined on  $\Lambda_L$ and on ${R}$, respectively.
Note that $f^{(L)}(x) \uparrow f(x)$ for $L\rightarrow\infty$. Then we obtain
\begin{equation}\label{Rxieta}
              \langle f_L, \xi \rangle = \langle f, \eta \rangle
           =   \langle f, T_R \, \xi \rangle \ ,
\end{equation}
see (\ref{T_D}).
{From} (\ref{gf0}), (\ref{gfD}), (\ref{Rxieta}) and (\ref{TR}), similar to arguments we used above it follows that
\[
        \int_{{\cal M}(R)}
       e^{-\langle f, \eta\rangle} \mathscr{M}^{(R)}_{\mu, L}
        (d\eta)
         = \int_{Q(\Lambda)}
         e^{-\langle f_L, \xi\rangle} \nu_{\mu, \Lambda}(d\xi)
\]
\[
      = \Det\big[1 + \sqrt{1-e^{-f_L}}Q_1K_{\Lambda_L}Q_1 \sqrt{1-e^{-f_L}}\big]^{-1}
\]
\[
       \times \Det[1 + \sqrt{1-e^{-f_L}}Q_0K_{\Lambda_L}Q_0\sqrt{1-e^{-f_L}}
              (1 + \sqrt{1-e^{-f_L}}Q_1K_{\Lambda_L}Q_1 \sqrt{1-e^{-f_L}})^{-1}]^{-1} \ .
\]

Again as above, the following estimate in the operator norm is obvious:
\begin{lem}\label{lem51}
\[
    \| \sqrt{1-e^{-f_L}}Q_1K_{\Lambda_L}Q_1\sqrt{1-e^{-f_L}}\|
       =O(L^{-1}).
\]
\end{lem}
We also have the following limit:
\begin{lem}\label{lem52}
\[
    \Tr [\sqrt{1-e^{-f_L}}Q_1K_{\Lambda_L}Q_1\sqrt{1-e^{-f_L}}]
    \xrightarrow[L\to \infty]{} K(x,x)\int_Rf(x)\, dx
\]
\end{lem}
{\sl Proof : } Similar to proof of Lemma \ref{lem32}, we get
\[
  \Tr[\sqrt{1-e^{-f_L}}Q_1K_{\Lambda_L}Q_1\sqrt{1-e^{-f_L}}]
\]
\begin{equation}\label{521}
        = \sum_{s \in \N^3, (s_1,s_3) \ne (1,1)}
       \frac{1}{L^4}
         \frac{\int_Rf^{(L)}(y)\, dy}
    {\exp\Big[\frac{\beta\hbar^2\pi^2}{2m}
      \sum_{j=1}^3\frac{s_j^2-1}{L_j^2}+\beta\Delta  \Big] - 1} + \widehat{R}_L \ ,
\end{equation}
where
\[
          \widehat{R}_L := \sum_{s \in \N^3, (s_2, s_3) \ne (1, 1)}
       \frac{1}{L^4}
         \frac{1}{\exp\Big[\frac{\beta\hbar^2\pi^2}{2m}
      \sum_{j=1}^3\frac{s_j^2-1}{L_j^2}+\beta\Delta  \Big] - 1}
\]
\[
      \times \int_Rf^{(L)}(y)\Big[
             - \sum_{j=1}^3(-1)^{s_j}\cos\Big(\frac{2\pi s_j}{L_j}Ly_j\Big)
             + \sum_{1\leqslant j<k \leqslant 3}(-1)^{s_j+s_k}
                \cos\Big(\frac{2\pi s_j}{L_j}Ly_j\Big)
               \cos\Big(\frac{2\pi s_k}{L_j}Ly_k\Big)
\]
\[             -  \prod_{j=1}^3(-1)^{s_j}\cos\Big(\frac{2\pi s_j}{L_j}Ly_j
              \Big) \Big]\, dy.
\]
In this case, we put
\[
     \Psi^{(R)}_L(p)
    := \begin{cases}
          \frac{1}{\exp\Big[\frac{\beta\hbar^2\pi^2}{2m}
          \sum_{j=1}^3\frac{s_j^2-1}{L_j^2}+\beta\Delta  \Big] - 1}
            & \; \mbox{ for } p_j
           \in  \Big( \frac{s_j-1}{L_j}, \frac{s_j}{L_j}\Big],
               \; j= 1,2,3, \; s\in \N^3,\, (s_2,s_3) \ne (1,1)         \\
          0 & \; \mbox{ for } p_2, p_3 \in
           \Big( 0, \frac{1}{L_3}\Big],
       \end{cases}
\]
and
\[
         a_L^{(R)}(p) := \int_Rf^{(L)}(y)\Big[
             - \sum_{j=1}^3(-1)^{s_j}\cos\Big(\frac{2\pi s_j}{L_j}Ly_j\Big)
             + \sum_{1\leqslant j<k \leqslant 3}(-1)^{s_j+s_k}
                \cos\Big(\frac{2\pi s_j}{L_j}Ly_j\Big)
               \cos\Big(\frac{2\pi s_k}{L_j}Ly_k\Big)
\]
\[
             -  \prod_{j=1}^3(-1)^{s_j}\cos\Big(\frac{2\pi s_j}{L_j}Ly_j\Big)
               \Big]\, dy
               \quad \mbox{ for } p_j
               =\Big( \frac{s_j-1}{L_j}, \frac{s_j}{L_j}\Big],
               \; j= 1,2,3, \; s\in \N^3.
\]
as before.
Then we have
\[
          |\Psi^{(R)}_L(p)| \leqslant \Phi(p) \equiv
          \frac{1}{e^{\beta\hbar^2\pi^2|p|^2/4m}-1},
          \quad \Phi \in L^1\big((0, \infty)^3\big),
\]
\[
      \lim_{L\to \infty}\Psi^{(R)}_L(p) = \frac{1}{e^{\beta\hbar^2\pi^2|p|^2/2m
       + \beta\Delta}-1} \quad a.e. ,
\]
and  $\lim_{L\to\infty}a_L^{(R)}(p) =  0 $.
Therefore by the dominated convergence theorem, we get
\[
      \lim_{L\to \infty} \widehat{R}_L = \lim_{L\to \infty}\int_{(0, \infty)^3}
        a_L^{(R)}(p)\Psi^{(R)}_L(p) \, dp = 0 \ ,
\]
and for the first term of the left hand side of eq.(\ref{521})
\begin{equation}
     \int_Rf^{(L)}(u)\, du \int_{[0, \infty)^3}
        \Psi^{(R)}_L(p)\, dp
      \longrightarrow \int_Rf(u)\, du\int_{[0, \infty)^3}
             \frac{dp}{e^{\beta\hbar^2\pi^2|p|^2/2m
       + \beta\Delta}-1} \ .
\end{equation}
\hfill $\square$
\begin{remark}\label{DetQ1}
   From the above two lemmata we obtain
\[
        \lim_{L\to\infty} \Det\big[1 + \sqrt{1-e^{-f_L}}Q_1K_{\Lambda_L}Q_1
         \sqrt{1-e^{-f_L}}\big]
       = e^{ K(x, x){\textstyle \int _R}f(x)\,dx} \ .
\]
\end{remark}
\begin{lem}\label{lem52}
\begin{eqnarray*}
&& \lim_{L\to \infty} \Det[1 + \sqrt{1-e^{-f_L}}Q_0K_{\Lambda_L}Q_0 \sqrt{1-e^{-f_L}}
(1 + \sqrt{1-e^{-f_L}}Q_1K_{\Lambda_L}Q_1 \sqrt{1-e^{-f_L}})^{-1}] \\
&& = \Det[1+ b \sqrt f I \otimes \cos\pi x_2(\cos\pi x_2, \otimes \cos\pi x_3(\cos\pi x_3, \sqrt f \ \cdot))] \\
&& = 1+ b \int_{R}f(y)\,\cos^2\pi y_2\cos^2\pi y_3\,dy  \ .
\end{eqnarray*}
Here, the Fredholm determinant in the first member is for
operators on $L^2(\Lambda_L)$, the second determinant is for operator on $L^2(R)$.
Here $I$ is the identity operator on $L^2(\R)$, which is considered as a component of the space
$L^2(R) = L^2(\R)\otimes L^2(-1/2, 1/2]) \otimes L^2(-1/2, 1/2])$.
\end{lem}
{\sl Proof:} Let $U_L^{(R)}: L^2(\Lambda_L) \to L^2(R_L)$ be unitary operator defined by $(U_Lg)(x):= L^{3/2}g(Lx)$ for
$x\in R_L, g \in L^2(\Lambda_L)$. Then, the corresponding integral kernels transform as
\[
        (U_L^{(R)}K\big(U_L^{(R)}\big)^{-1})(x, y)=
        L^3K(Lx, Ly)
\]
for $x,y \in R_L$. Then it follows that by (\ref{X1/X})
\[
    L^{-3}(U_L^{(R)}Q_0K_{\Lambda_L}Q_0
       \big(U_L^{(R)}\big)^{-1})(x, y)
     = Q_0K_{\Lambda}Q_0 (Lx, Ly)
\]
\begin{equation}
       =  \sum_{s_1=1}^{\infty}
       \frac{8}{L^4}
          \frac{
    \cos(\pi x_2)\cos(\pi y_2)\cos(\pi x_3)\cos(\pi y_3)}
     {\exp\Big[\frac{\beta\hbar^2\pi^2}{2m}
      \frac{s_1^2-1}{L^4}+\beta\Delta  \Big] - 1}
      \sin(\frac{\pi s_1}{2}+\frac{\pi s_1}{L^2}Lx)
      \sin(\frac{\pi s_1}{2}+\frac{\pi s_1}{L^2}Lx)
\end{equation}
\[
        \longrightarrow \frac{8m}{\beta\hbar^2\pi^2}
        \sum_{s_1=1}^{\infty}
     \frac{\cos0 - \cos\pi s_1}
          {s_1^2 + \frac{2mL^4\Delta|_{L\to \infty}}
           {\hbar^2\pi^2} -1}
        \cos(\pi x_2)\cos(\pi y_2)\cos(\pi x_3)\cos(\pi y_3),
\]
\[
         = \tilde \kappa \ \cos(\pi x_2)\cos(\pi y_2)
         \cos(\pi x_3)\cos(\pi y_3) \ ,
\]
locally uniformly in $x, y$. Recall that here we used
\[
    \tilde\kappa := \frac{16m}{\beta\hbar^2\pi^2}
    \varphi\Big(\frac{2mL^4\Delta|_{L\to \infty}}
           {\hbar^2\pi^2}\Big) \ ,
\]
and definition (\ref{phix}).

Then, we obtain that the $\|\cdot\|_1$-norm limit
\[
       U^{(R)}_L\sqrt{1-e^{-f_L}}Q_0K_{\Lambda}Q_0\sqrt{1-e^{-f_L}}\big(U^{(R)}_L\big)^{-1}
     = \sqrt{f^{(L)}}L^{-3}U^{(R)}_LQ_0K_{\Lambda_L}Q_0
   \big(U^{(R)}_L\big)^{-1}\sqrt{f^{(L)}}
\]
\[
          \longrightarrow \tilde \kappa \sqrt f  I
          \otimes \cos(\pi x_2)(\cos(\pi x_2),
          \otimes\cos(\pi x_3)(\cos(\pi x_3),
          \sqrt f \ \cdot )) \ ,
\]
exists. Thus, we get the lemma.  \hfill$\square$

Remark \ref{DetQ1} and Lemma \ref{lem52} yield Theorem\,\ref{thmE}.

\subsection{Proof of Theorem\,\ref{thmF}}
For non-negative continuous function $f$ on $ I =[-1/2,  1/2]$, we define
\[
    f^{(L)}(u) := L^4\big(1-e^{-f(u)/L^4}\big) \ ,
\]
and
\[
    f_L(x) := L^{-4}f\big( x_1/L^2\big) \ ,
\]
for $x = (x_1, x_2, x_3)$.
The functions $f_L, f^{(L)}$ are defined on  $\Lambda_L$ and on $I$, respectively.
Note that $f^{(L)} \uparrow f $ holds point-wise for $L\rightarrow\infty$.

By (\ref{T_D}) we have
\begin{equation}\label{Ixieta}
       \langle f_L, \xi \rangle =
          \langle f,  T_I \ \xi \rangle \ ,
\end{equation}
and  (\ref{gf0}), (\ref{gfD}) and (\ref{MIT}) yield
\[
        \int_{{\cal M}(I)}
       e^{-\langle f, \eta\rangle} \mathscr{M}^{(I)}_{\mu, L}
        (d\eta)
         = \int_{Q(\Lambda)}
         e^{-\langle f_L, \xi\rangle} \nu_{\mu, \Lambda}(d\xi)
\]
\[
      = \Det\big[1 + \sqrt{1-e^{-f_L}}Q_1K_{\Lambda_L}Q_1 \sqrt{1-e^{-f_L}}\big]^{-1}
\]
\[
       \times \Det[1 + \sqrt{1-e^{-f_L}}Q_0K_{\Lambda_L}Q_0\sqrt{1-e^{-f_L}}
              (1 + \sqrt{1-e^{-f_L}}Q_1K_{\Lambda_L}Q_1 \sqrt{1-e^{-f_L}})^{-1}]^{-1},
\]
similar to what we established above.

Again the following estimate in operator norm is obvious:
\begin{lem}\label{lem61}
\[
    \| \sqrt{1-e^{-f_L}}Q_1K_{\Lambda_L}Q_1\sqrt{1-e^{-f_L}}\|
       =O(L^{-2}).
\]
\end{lem}
We also have the following limit:
\begin{lem}\label{lem62}
\[
    \Tr [\sqrt{1-e^{-f_L}}Q_1K_{\Lambda_L}Q_1\sqrt{1-e^{-f_L}}]
    \xrightarrow[L\to \infty]{} K(x,x)\int_I f(u)\, du
\]
\end{lem}
{\sl Proof:} Similar to the proof of Lemma \ref{lem32}, we get
\[
  \Tr[\sqrt{1-e^{-f_L}}Q_1K_{\Lambda_L}Q_1\sqrt{1-e^{-f_L}}]
\]
\begin{equation}\label{621}
        = \sum_{s \in \N^3, (s_2,s_3) \ne (1,1)}
       \frac{1}{L^4}
         \frac{\int_I f^{(L)}(u)\, du}
    {\exp\Big[\frac{\beta\hbar^2\pi^2}{2m}
      \sum_{j=1}^3\frac{s_j^2-1}{L_j^2}+\beta\Delta  \Big] - 1} + \widetilde{R}_L \ ,
\end{equation}
where
\[
          \widetilde{R}_L := \sum_{s \in \N^3, (s_2, s_3) \ne (1, 1)}
       \frac{1}{L^4}
         \frac{1}{\exp\Big[\frac{\beta\hbar^2\pi^2}{2m}
      \sum_{j=1}^3\frac{s_j^2-1}{L_j^2}+\beta\Delta  \Big] - 1}
\]
\[
      \times \int_\mathcal{I} f^{(L)}(y)\Big[
             - \sum_{j=1}^3(-1)^{s_j}\cos\Big(\frac{2\pi s_j}{L_j}Ly_j\Big)
             + \sum_{1\leqslant j<k \leqslant 3}(-1)^{s_j+s_k}
                \cos\Big(\frac{2\pi s_j}{L_j}Ly_j\Big)
               \cos\Big(\frac{2\pi s_k}{L_j}Ly_k\Big)
\]
\[             -  \prod_{j=1}^3(-1)^{s_j}\cos\Big(\frac{2\pi s_j}{L_j}Ly_j
              \Big) \Big]\, dy.
\]
In this case, we put
\[
     \Psi^{(\mathcal{\mathcal{I}})}_L(p)
    := \begin{cases}
          \frac{1}{\exp\Big[\frac{\beta\hbar^2\pi^2}{2m}
          \sum_{j=1}^3\frac{s_j^2-1}{L_j^2}+\beta\Delta  \Big] - 1}
            & \; \mbox{ for } p_j
           \in  \Big( \frac{s_j-1}{L_j}, \frac{s_j}{L_j}\Big],
               \; j= 1,2,3, \; s\in \N^3,\, (s_2,s_3) \ne (1,1)         \\
          0 & \; \mbox{ for } p_2, p_3 \in
           \Big( 0, \frac{1}{L_3}\Big]
       \end{cases}
\]
and
\[
    a_L^{(I)}(p) = \int_If^{(L)}(y)\Big[
             - \sum_{j=1}^3(-1)^{s_j}\cos\Big(\frac{2\pi s_j}{L_j}Ly_j\Big)
             + \sum_{1\leqslant j<k \leqslant 3}(-1)^{s_j+s_k}
                \cos\Big(\frac{2\pi s_j}{L_j}Ly_j\Big)
               \cos\Big(\frac{2\pi s_k}{L_j}Ly_k\Big)
\]
\[             -  \prod_{j=1}^3(-1)^{s_j}\cos\Big(\frac{2\pi s_j}{L_j}Ly_j
              \Big) \Big]\, dy
       \qquad \mbox{ for } \quad  p_j
           \in  \Big( \frac{s_j-1}{L_j}, \frac{s_j}{L_j}\Big],
               \; j= 1,2,3, \; s\in \N^3  .
\]
Then we have
\[
          |\Psi^{(I)}_L(p)| \leqslant \Phi(p) \equiv
          \frac{1}{e^{\beta\hbar^2\pi^2|p|^2/4m}-1},
          \quad \Phi \in L^1\big((0, \infty)^3\big),
\]
\[
      \lim_{L\to \infty}\Psi_L(p) = \frac{1}{e^{\beta\hbar^2\pi^2|p|^2/2m
       + \beta\Delta}-1} \quad a.e..
\]
and $ \lim_{L\to \infty}a_L^{(I)}(p) =0 $.
The dominated convergence theorem yields
\[
      \lim_{L\to \infty} \widetilde{R}_L = \lim_{L\to \infty}\int_{(0, \infty)^3}
        a_L(p)\Psi_L(p) \, dp = 0
\]
and
\[
     \int_I f^{(L)}(u)\, du \int_{[0, \infty)^3}
        \Psi^{(I)}_L(p)\, dp
      \longrightarrow \int_I f(u)\, du\int_{[0, \infty)^3}
             \frac{dp}{e^{\beta\hbar^2\pi^2|p|^2/2m
       + \beta\Delta}-1}
\]
for the first term of the left hand side of eq.(\ref{621}).
\hfill $\square$
\begin{remark}\label{DetQ2}
The above two lemmata imply:
\[
        \lim_{L\to\infty} \Det\big[1 + \sqrt{1-e^{-f_L}}Q_1K_{\Lambda_L}Q_1
         \sqrt{1-e^{-f_L}}\big]
       = e^{ K(x, x){\textstyle \int _I}f(u)\,du} \ .
\]
\end{remark}
\begin{lem}\label{lem62}
\begin{eqnarray*}
&&\lim_{L\to \infty} \Det[1 + \sqrt{1-e^{-f_L}}Q_0K_{\Lambda_L}Q_0
\sqrt{1-e^{-f_L}}(1 + \sqrt{1-e^{-f_L}}Q_1K_{\Lambda_L}Q_1 \sqrt{1-e^{-f_L}})^{-1}]  \\
&&= \Det[1+\sqrt f R \sqrt f \otimes \cos\pi x_2(\cos\pi x_2, \otimes \cos\pi x_3(\cos\pi x_3, \ \cdot))]\\
&& = \Det[ 1 + 4^{-1}\sqrt f R \sqrt f] \ .
\end{eqnarray*}
Here, the Fredholm determinant in the first member is for operators on $L^2(\Lambda_L)$, the second determinant
stays for operators on $L^2(I^3)$ and the third one for operators on $L^2(I)$ , see (\ref{Ruv}).
\end{lem}
{\sl Proof:} Let $U_L: L^2(\Lambda_L) \to L^2(I^3)$ be unitary operator $(U_Lg)(x):= L^{2}g(L^2x_1, Lx_2, Lx_3)$
for $x\in I^3, g \in L^2(\Lambda_L)$. Then integral kernels transform as
\[
        (U_LKU_L^{-1})(x, y)=
        L^4K((L^2x_1, Lx_2, Lx_3),
        (L^2y_1, Ly_2, Ly_3)) \ ,
\]
for $x, y \in I^3$. It follows that
\[
    L^{-4}(U_LQ_0K_{\Lambda_L}Q_0U_L^{-1})(x, y)
     = Q_0K_{\Lambda}Q_0 ((L^2x_1, Lx_2, Lx_3),
        (L^2y_1, Ly_2, Ly_3))
\]
\begin{equation}
       =  \sum_{s_1=1}^{\infty}
       \frac{8}{L^4}
          \frac{
       \sin(\pi s_1(2^{-1}+x_1) \sin(\pi s_1(2^{-1}+y_1)
    \cos(\pi x_2)\cos(\pi y_2)\cos(\pi x_3)\cos(\pi y_3)}
     {\exp\Big[\frac{\beta\hbar^2\pi^2}{2m}
      \frac{s_1^2-1}{L^4}+\beta\Delta  \Big] - 1}
\end{equation}
\[
        \longrightarrow \frac{8m}{\beta\hbar^2\pi^2}
        \sum_{s_1=1}^{\infty}
     \frac{\cos[\pi s_1(x_1-y_1)] - \cos[\pi s_1(1+x_1+y_1)]}
          {s_1^2 + \frac{2mL^4\Delta|_{L\to \infty}}
           {\hbar^2\pi^2} -1}
        \cos(\pi x_2)\cos(\pi y_2)\cos(\pi x_3)\cos(\pi y_3),
\]
\[
         = R(x_1, y_1)\cos(\pi x_2)\cos(\pi y_2)
         \cos(\pi x_3)\cos(\pi y_3)
\]
uniformly in $x, y$, see  (\ref{Ruv}) and (\ref{X1/X}).

Then, we get the trace-norm limit:
\[
       U^{(I)}_L\sqrt{1-e^{-f_L}}Q_0K_{\Lambda}Q_0\sqrt{1-e^{-f_L}}\big(U^{(I)}_L\big)^{-1}
     = \sqrt{f^{(L)}}L^{-4}U^{(I)}_LQ_0K_{\Lambda_L}Q_0
   \big(U^{(I)}_L\big)^{-1}\sqrt{f^{(L)}}
\]
\[
          \longrightarrow \sqrt f R \sqrt f
          \otimes \cos(\pi x_2)(\cos(\pi x_2),
          \otimes\cos(\pi x_3)(\cos(\pi x_3) \ , \ \cdot )) \ ,
\]
which proves the lemma.  \hfill$\square$

Remark \ref{DetQ2} and Lemma \ref{lem62} yield
Theorem\,\ref{thmF}.

\bigskip

\noindent \textbf{Acknowledgements}

\bigskip

\noindent H.T. thanks  JSPS for financial support under the Grant-in-Aid for Scientific Research (C) 20540162.
H.T. is also thankful the hospitality of CPT-Luminy, since essential part of this paper was finished there
during his short visit in 2011.

\appendix
\section{Miscellaneous formulae }
In the appendix, the formulae used in the argument of this article
are shown.
Here, $A$ is supposed to be a positive constant independent
of $L$ or $L_1, L_2, L_3$, while a positive constant $B$ to depend
on  $L$ or $L_1, L_2, L_3$ smoothly.
We use the notation $s=(s_1,s_2,s_3) $ for $s \in \N^3$ and
$s=(s_1,s_2) $ for $s \in \N^2$, and so on.

\noindent { \large 1$^{\circ}$}
\begin{equation}
    \sum_{s=(s_1,s_2) \in \N^2 \atop s \ne (1,1)} \frac{1}{L^2}
   \frac{1}{\exp\Big[A\Big(\frac{s_1^2-1}{L^2} +
       \frac{s^2_2-1}{L^2} \Big)   +B\Big] -1} =
    \frac{\pi}{4A}\log(L^2\wedge B^{-1}) + O(1). \quad
\label{A1}
\end{equation}

\begin{equation}
   \sum_{s=(s_1,s_2)\in \N^2 \atop s\ne(1,1), s_1, s_2: odd }
    \frac{1}{L^2}
    \frac{1}{\exp\Big[A\Big(\frac{s_1^2-1}{L^2} +
    \frac{s^2_2-1}{L^2} \Big)  +B\Big] -1}
     = \frac{\pi}{16A}\log(L^2\wedge B^{-1}) + O(1).
\label{A2}
\end{equation}

\begin{equation}
   \sum_{s \in \N, s:odd, s>1 } \frac{1}{L}
    \frac{1}{\exp\Big[A\frac{s^2-1}{L^2} +B\Big] -1}
     = \frac{L}{A}\varphi\Big(\frac{L^2B}{A}\Big) + O(1),
\label{A21}
\end{equation}
where $\varphi$ is defined in (\ref{phix}).

\noindent { \large 2$^{\circ}$}

Suppose $L_1 \geqslant L_2 \geqslant L_3$. The following convergence holds
uniformly in $x, y$ on any compact set as $L_1, L_2, L_3\to\infty$.
\begin{equation}
             \sum_{s\in \N^3 \atop s_3 \ne 1}
            \frac{\phi_{s,\Lambda}(x)\overline{\phi_{s,\Lambda}(y)}}
              {\exp\Big[A\Big(\frac{s_1^2-1}{L_1^2}+ \frac{s_2^2-1}{L_2^2}
           + \frac{s_3^2-1}{L_3^2}\Big) + B\Big] -1}
         \longrightarrow \displaystyle\int_{\R^3}
            \frac{e^{2\pi ip\cdot(x-y)}dp}{e^{4A|p|^2+B(L=\infty)}-1}.
\label{A3}
\end{equation}
If $L_1 \geqslant L_2 = L_3$. The following convergence holds
uniformly in $x, y$ on any compact set as $L_1, L_2, L_3\to\infty$.
\begin{equation}
         \sum_{s\in \N^3 \atop (s_2,s_3) \ne (1,1)}
          \frac{\phi_{s,\Lambda}(x)\overline{\phi_{s,\Lambda}(y)}}
           {\exp\Big[A\Big(\frac{s_1^2-1}{L_1^2}+ \frac{s_2^2-1}{L_2^2}
           + \frac{s_3^2-1}{L_3^2}\Big) + B\Big] -1}
           \longrightarrow \displaystyle \int_{\R^3}
            \frac{e^{2\pi ip\cdot(x-y)}dp}{e^{4A|p|^2+B(L=\infty)}-1}.
\label{A31}
\end{equation}
Suppose $L_1 \geqslant L_2 \geqslant L_3 $ and ${\bf 1} = (1, 1, 1)$.
Then,  we have
\[
\sum_{s\in \N^3 \atop s_3 \ne 1} \frac{1}{L_1L_2L_3}
   \frac{1}{\exp\Big[A\Big(\frac{s_1^2-1}{L_1^2}+ \frac{s_2^2-1}{L_2^2}
    + \frac{s_3^2-1}{L_3^2}\Big) + B\Big] -1}
\]
\begin{equation}
    \longrightarrow
      \frac{1}{8}\int_{\R^3}\frac{dp}{e^{A|p|^2+B(L=\infty)}-1}
\label{A4}
\end{equation}
as $L_1, L_2,  L_3 \to \infty$.

\noindent { \large 3$^{\circ}$}

Suppose, $L_1 \geqslant L_2$. Then,
\[
    \sum_{s\in \N^2 \atop s\ne (1,1)} \frac{1}{L_1L_2}
   \frac{1}{\exp\Big[A\Big(\frac{s_1^2-1}{L_1^2}+
     \frac{s_2^2-1}{L_2^2}\Big) + B \Big] -1}
\]
\begin{equation}
    = O\Big(\frac{L_1}{L_2}\wedge\frac{1}{L_2\sqrt B}\Big) +
   O\big(\log(L_2\wedge B^{-1/2})\big)
\label{A7}
\end{equation}
holds for large $L_1, L_2$, and $L_3$.

The following asymptotics also hold for large $L$:
\begin{eqnarray}
    \sum_{s\in \N^2} \frac{1}{L^2}\frac{s_1^2+s_2^2}{L^2}
   \frac{1}{\exp\Big[A\frac{s_1^2-1+s_2^2-1}{L^2} +B\Big] -1} =
    O(1) + O(L^{-4}B^{-1}).
\label{A5}  \\
\sum_{s\in \N^2} \frac{1}{L^2}\Big(\frac{s_1^2+s_2^2}{L^2}\Big)^{1/4}
   \frac{1}{\exp\Big[A\frac{s_1^2-1+s_2^2-1}{L^2} +B\Big] -1} =
    O(1) + O(L^{-4}B^{-1}).
\label{A501}  \\
\sum_{s\in \N} \frac{1}{L}\frac{s^2}{L^2}
   \bigg(\exp\Big[A\frac{s^2-1}{L^2} +B\Big] -1\bigg)^{-1} =
    O(1) + O(L^{-5/2}B^{-1}).
\label{A51}
\end{eqnarray}
\begin{equation}
    \sum_{s=2}^{\infty} \frac{1}{L}
   \bigg(\exp\Big[A\frac{s^2-1}{L^2} +B\Big] -1\bigg)^{-1} =
    O(L\wedge B^{-1/2}).
\label{A6}
\end{equation}

\noindent{\bf Proof} of (\ref{A1}): Put
\[
    I_L =  \sum_{s\in \N^2 \atop s\ne (1,1)} \frac{1}{L^2}
   \frac{1}{\exp\Big[A\frac{s_1^2-1+s_2^2-1}{L^2} +B\Big] -1}.
\]
Because of
\[
     p_j^2 \leqslant \frac{s_j^2}{L^2} \leqslant
      \Big(p_j+\frac{1}{L}\Big)^2
    \qquad \mbox{ for } \quad p_j \in \Big(\frac{s_j-1}{L},
    \frac{s_j}{L}\Big] \qquad ( s_j \in \N, j=1,2 ),
\]
we get
\[
       L.B. := \int _{(0, \infty)^2-(0, 1/L]^2}\frac{dp}
       {\exp[A((p_1 +1/L)^2+(p_2 +1/L)^2-2/L^2) +B]-1}
        \leqslant I_L
\]
\[
    \leqslant    \frac{1}{L^2} \frac{8}{\exp[3A/L^2 +B]-1} +
     \int _{(0, \infty)^2-(0, 3/L]^2}\frac{dp}
       {\exp[A(|p|^2-2/L^2) +B]-1} =: U.B. \ .
\]
For the \textit{lower bound} (L.B.), we have
\[
     L.B. = \int _{(1/L, \infty)^2-(1/L, 2/L]^2}
           \frac{dp}{\exp[A(|p|^2-2/L^2) +B]-1}
           \geqslant \int_{\{|p| \geqslant 2\sqrt 2/L,
        \, p_1, p_2 > 1/L \}}
\]
\[
     = \int_{2\sqrt2/L}^{\infty}pdp \int_{\sin^{-1}1/pL}
      ^{\pi/2- {\sin^{-1}1/pL}}d\theta
      \frac{1}{\exp[A(p^2-2/L^2) +B]-1}
\]
\[
      \geqslant \frac{\pi}{2}\int_{2\sqrt2/L}^{\infty}\frac{pdp}
      {\exp[A(p^2-2/L^2) +B]-1}
      - 2\int_{2\sqrt2/L}^{\infty}\frac{p\sin^{-1}(1/pL)\, dp}
       {\exp[A(p^2-2/L^2) +B]-1}.
\]
Here the first term is calculated as
\[
     \frac{\pi}{2} \frac{1}{2A}\log \frac{1}{1-e^{-6A/L^2 -B}}
     = \frac{\pi}{4A}\log \Big(L^2 \wedge \frac{1}{B}\Big) + O(1).
\]
For the second term, we use $ \sin^{-1}x \leqslant \pi x/2 $ to get
\[
    \mbox{the 2nd term} \geqslant -\frac{\pi}{L}
    \int_{2\sqrt2/L}^{\infty}\frac{dp}
       {\exp[A(p^2-2/L^2) +B]-1}
\]
\[
      \geqslant -\frac{\pi}{AL}
    \int_{2\sqrt2/L}^{\infty}\frac{dp}
       {p^2-2/L^2} = O(1).
\]
For the \textit{upper bound} (U.B.), we enlarge the integral domain and perform the
angle integral to get
\[
     U.B. \leqslant \frac{\pi}{2}\int_{3/L}^{\infty}\frac{pdp}
      {\exp[A(p^2-2/L^2) +B]-1} + \frac{1}{L^2} \frac{8}{\exp[3A/L^2 +B]-1}
\]
\begin{equation}
    \frac{\pi}{4A}\log \frac{1}{1-e^{-7A/L^2 -B}} + O(1)
    = \frac{\pi}{4A}\log \Big(L^2\wedge\frac{1}{B}\Big) + O(1).
\tag*{$\square$}
\end{equation}

\noindent{\bf Proof } of (\ref{A2}) : For
\[
      S_{n,m} = \frac{1}{L^2} \frac{1}
   {\exp\Big[A\frac{(2n-1)^2-1+(2m-1)^2-1}{L^2} +B\Big] -1}.
\]
the inequalities
\[
     \Big(2p_1-\frac{1}{L}\Big)^2 \leqslant \frac{(2n-1)^2}{L^2}
        \leqslant \Big(2p_1+\frac{1}{L}\Big)^2
    \qquad \mbox{ for } \quad p_1 \in \Big(\frac{n-1}{L}, \frac{n}{L}\Big],
\]
\[
     \Big(2p_2-\frac{1}{L}\Big)^2 \leqslant \frac{(2m-1)^2}{L^2}
     \leqslant \Big(2p_2+\frac{1}{L}\Big)^2
    \qquad \mbox{ for } \quad p_2 \in \Big(\frac{m-1}{L}, \frac{m}{L}\Big],
\]
hold for $ n, m = 2, 3, \cdots$. The upper bounds of the above inequalities
also hold for $ n= 1 $ or $ m = 1$. Thereby, we have
\[
     \int _{(0, \infty)^2-(0, 1/L]^2}\frac{dp}
       {\exp[A((2p_1 +1/L)^2+(2p_2 +1/L)^2-2/L^2) +B]-1}
\]
\[
        \leqslant \sum_{n, m\in \N,\atop (n,m)\ne (1,1)}S_{n,m}
        \leqslant 3S_{2,1} + 4\sum_{n=3}^{\infty}S_{n,1}
         + \sum_{n,m = 3}^{\infty}S_{n,m} .
\]
Calculation of these bounds can be performed in a similar way as
in the proof of (\ref{A1}). \hfill $\square$

\noindent{\bf Proof} of (\ref{A21}): Using
\begin{equation}
      0 \leqslant \frac{1}{X} - \frac{1}{e^X-1}
      \leqslant X \wedge \frac{1}{X},
\label{X1/X}
\end{equation}
it is obvious to get
\[
    \bigg|\sum_{n \in \N } \frac{1}{L}
    \frac{1}{\exp\Big[A\frac{(2n-1)^2-1}{L^2} +B\Big] -1}
     - \frac{L}{A}\varphi\Big(\frac{L^2B}{A}\Big) \bigg|
\]
\begin{equation}
     = \sum_{n \in \N } \frac{1}{L}
    \bigg|\frac{1}{\exp\Big[A\frac{(2n-1)^2-1}{L^2} +B\Big] -1}
      \; - \; \frac{1}{A\frac{(2n-1)^2-1}{L^2} +B} \bigg|
     =  O(1).
\tag*{$\square$}
\end{equation}

\bigskip

\noindent{\bf Proof } of (\ref{A3}): First,  note that the inequality
\begin{equation}
     \frac{(2n-1)^2 -1 }{L_1^2} + \frac{(2m-1)^2 -1}{L_2^2}
      + \frac{(2l)^2 -1 }{L_3^2}
     \geqslant p_1^2 +p_2^2 +p_3^2 =|p|^2
\label{nml1}
\end{equation}
holds for
\[
       p_1\in\Big(\frac{n-1}{L_1}, \frac{n}{L_1}\Big], \quad
       p_2\in\Big(\frac{m-1}{L_2}, \frac{m}{L_2}\Big], \quad
       p_3\in\Big(\frac{l-1}{L_3}, \frac{l}{L_3}\Big], \quad
       (n,m,l \in \N),
\]
and inequality
\begin{equation}
     \frac{(2n-1)^2 -1}{L_1^2} + \frac{(2m-1)^2-1}{L_2^2}
        + \frac{(2l-1)^2 -1}{L_3^2}
     \geqslant p_1^2 +p_2^2 +p_3^2 =|p|^2
\label{nml2}
\end{equation}
holds for
\[
       p_1\in\Big(\frac{n-1}{L_1}, \frac{n}{L_1}\Big], \quad
       p_2\in\Big(\frac{m-1}{L_2}, \frac{m}{L_2}\Big], \quad
       p_3\in\Big(\frac{l-1}{L_3}, \frac{l}{L_3}\Big], \quad
       (n,m,l \in \N, l\ne 1),
\]
since $L_1 \geqslant L_2 \geqslant L_3$.
Now set
\[
    \Psi_L^{oeo}(p;x,y) :=
              \Big[\sin\Big(\frac{(2n-1)\pi}{2}+\frac{(2n-1)\pi x_1}{L_1}\Big)
              \sin\Big(\frac{(2n-1)\pi}{2}+\frac{(2n-1)\pi y_1}{L_1}\Big)
\]
\[
              \times\sin\Big(\frac{2m\pi}{2}+\frac{2m\pi x_2}{L_2}\Big)
               \sin\Big(\frac{2m\pi}{2}+\frac{2m\pi y_2}{L_2}\Big)
\]
\[
              \times\sin\Big(\frac{(2l-1)\pi}{2}+\frac{(2l-1)\pi x_3}{L_1}\Big)
              \sin\Big(\frac{(2l-1)\pi}{2}+\frac{(2l-1)\pi y_3}{L_1}\Big)\Big]
\]
\[
              \bigg/\Big[\exp\Big[A\Big(\frac{(2n-1)^2-1}{L_1^2}
              + \frac{(2m)^2-1}{L_2^2}
              + \frac{(2l-1)^2-1}{L_3^2}\Big) +B\Big]-1\Big]
\]
if
\[
      \hphantom{\sin}     p_1\in\Big(\frac{n-1}{L_1}, \frac{n}{L_1}\Big], \quad
       p_2\in\Big(\frac{m-1}{L_2}, \frac{m}{L_2}\Big], \quad
       p_3\in\Big(\frac{l-1}{L_3}, \frac{l}{L_3}\Big], \quad
       (n,m,l \in \N, l\ne 1).
\]
Otherwise,  $  \Psi_L^{oeo}(p;x,y) =  0  $.
Similarly we can also define $\Psi_L^{eee},  \Psi_L^{eeo}$ etc.
Put
\[
      \Phi(p) =\frac{1}{e^{A|p|^2}-1}.
\]
Then,
\[
    \Phi \in L^1\big((0, \infty)^3\big) \quad \mbox{and} \quad
        |\Psi_L^{oeo}(p;x,y)| \leqslant \Phi(p)
\]
hold. Due to the dominated convergence theorem, we get
\begin{equation}
         \int_{(0, \infty)^3} \Psi_L^{oeo}(p;x,y) \, dp \longrightarrow
         \int_{(0, \infty)^3} \lim_{L_1,L_2,L_3 \to \infty}
      \Psi_L^{oeo}(p;x,y) \, dp.
\label{unifconv}
\end{equation}
In the same way we also get the convergence of the integrals of
$ \Psi_L^{eeo}(p;x,y) , \; \Psi_L^{ooe}(p;x,y) $ and so on, and finally
\[
      \sum_{s\in \N^3 \atop s_3 \ne 1}
            \frac{\phi_{s,\Lambda}(x)\overline{\phi_{s,\Lambda}(y)}}
              {\exp\Big[A\Big(\frac{s_1^2-1}{L_1^2}+ \frac{s_2^2-1}{L_2^2}
           + \frac{s_3^2-1}{L_3^2}\Big) + B\Big] -1}
\]
\[
      = 8\int_{(0, \infty)^3} \big( \Psi_L^{eee}(p;x,y) +
       \Psi_L^{eeo}(p;x,y) + \cdots + \Psi_L^{ooo}(p;x,y)\big) \, dp
       \longrightarrow
       \int_{\R^3}\frac{e^{2\pi ip\cdot(x-y)}}
         {e^{4A|p|^2+B(L=\infty)}-1}\,dp.
\]
Here, note that the convergence in (\ref{unifconv}) is uniform in $x, y$ on compact set.
This follows from the following obvious observations:

\noindent{\sl Suppose that $F, F_1, F_2, \cdots \in L^1(\R^3)$ satisfy
        $ \| F_n -F\|_1 \to 0$,
and that $f(p,x), f_1(p,x) ,f(p,x), \cdots $ are bounded functions of
$p,x \in \R^3$ satisfying $\| f_n -f\|_{\infty} \to 0$.}

\noindent {\sl Then, $\int F_n(p)f_n(p,x)\,dp \to \int F(p)f(p,x)\,dp$ \; holds
uniformly in $x$. \hfill $\square$}

\bigskip

\noindent{\bf Proof } of (\ref{A31}) is almost the same as the proof of
(\ref{A3}).

\noindent{\bf Proof } of (\ref{A4}) is the same as the proof of the limit of the integral for
$\Psi_L$ in Lemma \ref{lem32}.

\noindent{\bf Proofs} of the other formulae are straightforward.

\newpage



\begin{thebibliography}{999999}


\bibitem[BZ]{BZ} M. Beau and V. A. Zagrebnov, The second critical density and
anisotropic generalised condensation, Cond. Mat. Phys. {\bf 13} (2010) 23003: 1-10.

\bibitem[BLL]{BLL} M. van den Berg, J.T. Lewis and M. Lunn, On the general theory of Bose-Einstein Condensation
and the state of the free bose gas,
Helv. Phys. Acta {\bf 59} (1986) 1289-1310.

\bibitem[BLP]{BLP} M. van den Berg, J.T. Lewis and J. Pul\'e, A general theory of Bose-Einstein Condensation,
Helv. Phys. Acta {\bf 59} (1986) 1273-1288.

\bibitem[BrRo]{BrRo} O. Bratteli and D. W. Robinson,
{\em Operator algebrasand quantum statistical mechanics II} (Springer-Verlag, Berlin, 1996).

\bibitem[BNZ]{BNZ} J.-B. Bru, B. Nachtergaele and V. A. Zagrebnov,
The Equilibrium States for a Model with Two Kinds of Bose Condensation,
J. Stat. Phys. {\bf 109} (2002) 143--176.

\bibitem[BruZ]{BruZ} J.-B. Bru and V. A. Zagrebnov, Large Deviations in the Superstable Weakly Imperfect Bose-Gas,
J. Stat. Phys. {\bf 133} (2008) 379--400.

\bibitem[DV]{DV}  D. J. Daley and D. Vere-Jones, \textit{An Introduction to
the Theory of Point Processes } (Springer-Verlag, Berlin, 1988).

\bibitem[DGPS]{DGPS} F. Dalfovo, S.Giorgini, L. P. Pitaevskii and S. Stringari,
Theory of Bose-Einstein condensation in trapped gases, Rev. Mod. Phys. \textbf{71}
(1999) 463--512.

\bibitem[vDK]{vDK} N.J. van Druten, W. Ketterle,  Two-step Condensation of the Ideal Gas in Highly Anisotropic
Traps, Phys. Rev. Lett. {\bf 79} (1997) 549--552.

\bibitem[E]{E} N. Eisenbaum, A Cox process involved in the Bose-Einstein
condensation, Annales Henri Poincar\'e \textbf{9}  (2008) 1123--1140.

\bibitem[F]{F91} W. Freudenberg, Characterization of states of infinite boson systems. II:
On the existence of the conditional reduced density matrix. Commun. Math. Phys. {\bf 137},
461--472(1991)

\bibitem[FF]{FF91} K.-H. Fichtner and W. Freudenberg, Characterization of
states of infinite boson systems. I: On the construction of states of boson
systems. Commun. Math. Phys. {\bf 137} (1991) 315--357.

\bibitem[GeoY]{GeoY} H.-O. Georgii and H. J. Yoo, Conditional Intensity and Gibbsianness
of Determinantal Point Process. J. Stat. Phys. {\bf 118} (2005) 55--84.

\bibitem[GA]{G} F. Gerbier, J.H. Thywissen, S. Richard, M. Hugbart, P.
Bouyer, A. Aspect, Momentum distribution and correlation function of
quasicondensates in elongated traps, Phys. Rev. A {\bf 67} (2003) 051602(R): 1-4.

\bibitem[HD]{H} Z. Hadzibabic, P. Kr\"uger, M. Cheneau, B. Battelier,
J. Dalibard, Berezinskii-Kosterlitz-Thouless crossover in a trapped
atomic gas, Nature, {\bf 441} (2006) 1118--1121.

\bibitem[KvD]{KvD} W. Ketterle, N.J. van Druten, Bose-Einstein condensation of a finite number of particles
trapped in one or three dimensions, Phys. Rev. A{\bf 54} (1996) 656--660.

\bibitem[LePu]{LePu} J. Lewis and J. V. Pul\`{e},  {The Equilibrium States of
the Free Boson Gas}, Commun. Math. Phys. \textbf{36} (1974) 1--18.

\bibitem[LSSY]{LSSY} E. H. Lieb, R. Seiringer, J. P. Solovej and J. Yngvason,
\textit{The Mathematics of the Bose Gas and its Condensation} (Birkhaeuser,
Basel, 2005).

\bibitem[Ly]{Ly} E. Lytvynov, Fermion and boson random point processes as
particle distributions of infinite free Fermi and Bose gases of finite
density, Rev. Math. Phys. {\bf 14} (2002) 1073--1098.

\bibitem[M75]{M75} O. Macchi, The coincidence approach to stochastic point
processes, Adv. Appl. Prob. {\bf 7} (1975) 83--122.

\bibitem[MaVe]{MaVe} J. Manuceau, A. Verbeure, Quasi-free states of the
CCR-algebra and Bogoliubov transformations, Commun.Math.Phys.
\textbf{9} (1977) 125--131.

\bibitem[MuSa]{MuSa} W.J. Mullin, A.R. Sakhel, Generalized Bose-Einstein Condensation, J. Low Temp. Phys.
\textbf{166} (2012) 125-150,  arXiv:1012.2850v2 [quant-ph] pp.1--19 (2011).

\bibitem[Pu]{Pu} J. V. Pul\'{e}, The free boson gas in a weak external potential,
J. Math. Phys. \textbf{24} (1983) 138--142.

\bibitem[PuZ]{PuZ} J. V. Pul\'{e}, V. A. Zagrebnov, The canonical perfect Bose gas in Casimir boxes,
J.Math.Phys. \textbf{45} (2004) 3565--3583.

\bibitem[RSIV]{RSIV} M. Reed and B. Simon,
\textit{Methods of Modern Mathematical Physics}, vol.IV: Analysis of Operatyors
(Academic Press, London, 1978).

\bibitem[Ru]{Ru} D. Ruelle, \textit{Statistical Mechanics. Rigorous Results},
(W.A.Benjamin, Inc., Amsterdam, 1969).

\bibitem[ST]{ST03} T. Shirai and Y. Takahashi, Random point fields
associated with certain Fredholm determinants I:
fermion, Poisson and boson point processes,
J. Funct. Anal. {\bf 205} (2003) 414--463.

\bibitem[Sh]{Sh} A. N. Shiryaev, \textit{Probability}, (Springer Verlag, Berlin, 1996)

\bibitem[deS-Z]{deS-Z} Ph. de Smedt and V.A. Zagrebnov, Van der
Waals Limit of an Interacting Bose Gas in a Weak External Field,
Phys. Rev. \textbf{A35} (1987) 4763--4769.

\bibitem[So]{So} A. Soshnikov, Determinantal random point processes,
Russ.Math.Surveys {\bf 55} (2000) 923--975.

\bibitem[T]{T} H. Tamura, Boson Gas Mean Field Model Trapped by Weak
Harmonic Potentials in Mesoscopic Scaling, RIMS
K$\hat {\rm o}$ky$\hat {\rm u}$roku Bessatsu, {\bf B21}, (2010), 163-181

\bibitem[TIa]{1} H. Tamura and K.R. Ito,  A Canonical Ensemble Approach
to the Fermion/Boson Random Point Processes and its Applications,
Commun. Math. Phys. {\bf 263} (2006) 353--380.

\bibitem[TIb]{2} H. Tamura and K.R. Ito,  A Random Point Field related to
Bose-Einstein Condensation, J. Funct. Anal. {\bf 243} (2007) 207--231.

\bibitem[TIc]{3} H. Tamura and K.R. Ito,  Random Point Fields for
Para-Particles of Any Order, J. Math. Phys. {\bf 48} 023301: 1-14 (2007).

\bibitem[TZa]{TZa} H. Tamura and V.A. Zagrebnov: Mean-field
interacting boson random point fields in weak harmonic traps,
J. Math. Phys. {\bf 50}, 023301: 1-28 (2009).

\bibitem[TZb]{TZb} H. Tamura and V.A. Zagrebnov: Large deviation
principle for noninteracting boson random point processes,
J. Math. Phys. {\bf 51}, 023528: 1-20 (2010).

\bibitem[VVZ]{VVZ} L. Vandevenne, A. Verbeure and V. A. Zagrebnov,
Equilibrium states for the Bose gas,  J.Math.Phys. \textbf{45} (2004) 1606-1622.

\bibitem[Ver]{Ver} A. F. Verbeure,
\textit{ Many-Body Boson Systems}. \ (Springer-Verlag, Heidelberg-London 2011).

\bibitem[Zag]{Zag} V.A. Zagrebnov,
{\it Topics in the Theory of Gibbs Semigroups.} \ Leuven Notes in Mathematical and Theoretical Physics.
Vol. 10 (Leuven University Press, Leuven 2002).

\bibitem[Zag2]{Zag2} V.A. Zagrebnov, \textit{Bose-Einstein Condensation in Presence of External Potentials},
in Proceedings of the 3rd Warsaw School of Statistical Physics
 27 June – 4 July 2009, Kazimierz Dolny, Poland.

\bibitem[ZB]{ZB} V.A. Zagrebnov and J.-B. Bru,
The Bogoliubov model of weakly imperfect Bose gas, Phys.
Rep. {\bf 350} (2001) 291--434 .

\bibitem[ZP]{ZP} V.A. Zagrebnov and Vl.V. Papoyan, The ensemble equivalenceproblemfor Bosesystems (Non-ideal Bose gas)
Theor.Math.Phys. {\bf 69} (1986) 1240--1253 .

\end{thebibliography}
\end{document}